\pdfoutput=1
\documentclass[%
 reprint,
 amsmath,amssymb,
 aps,prb
]{revtex4-1}

\usepackage{graphicx}
\usepackage{dcolumn}
\usepackage{bm}
\usepackage{color}

\bmdefine{\boldi}{i}
\bmdefine{\boldj}{j}
\bmdefine{\boldl}{l}
\bmdefine{\boldS}{S}
\bmdefine{\boldq}{q}
\bmdefine{\boldzero}{0}
\bmdefine{\bolddelta}{\delta}

\begin{document}

\title{
  Controlling stability of Bose-Einstein condensation of interacting magnons
  in an antiferromagnet by an external magnetic field
}

\author{Naoya Arakawa}
\email{naoya.arakawa@sci.toho-u.ac.jp} 
\affiliation{
Department of Physics, Toho University, 
Funabashi, Chiba, 274-8510, Japan}

\date{\today}

\begin{abstract}
  We propose a mechanism for destabilizing or stabilizing 
  Bose-Einstein condensation (BEC) of interacting magnons in an antiferromagnet.
  We study how the interaction between magnons affects
  the magnon BEC in a two-sublattice antiferromagnet without and with
  an external magnetic field.
  We show that
  in the absence of the magnetic field 
  the magnon BEC is destabilized by
  the attractive interband interaction, which is larger than
  the repulsive intraband interaction.
  We also show that
  in the presence of the magnetic field 
  the magnon BEC is stabilized
  only if the magnetic field is large enough to
  make the intraband interaction larger than the interband interaction.
  Our results provide the first step for understanding
  the interaction effects on the magnon BEC in antiferromagnets
  and may open interaction physics of multicomponent BEC of magnons in magnets.

\end{abstract}
\maketitle

\section{Introduction}

Stability of Bose-Einstein condensation (BEC) depends on
the sign of the interaction between bosons~\cite{Leggett-review}.
BEC is the phenomenon that  
many bosons occupy the lowest-energy state~\cite{Einstein,BEC-text}.
This can be understood in terms of noninteracting bosons in principle. 
However, 
if the interaction is attractive, 
the BEC becomes unstable;
if the interaction is repulsive, the BEC remains stable~\cite{FW}. 
Thus
whether the interaction is attractive or repulsive
is vital for understanding the stability of the BEC.

Nevertheless,
it is unclear how
the interaction between magnons affects 
the BEC of magnons in an antiferromagnet.  
Since magnons are bosonic quasiparticles of a magnetically ordered system,
the BEC of magnons is expected to occur
in magnets~\cite{MagBEC-JETP,MagBEC-Nature,MagBEC-He,MagBEC-AF1,MagBEC-AF2,MagBEC-AF3}. 
In addition,
theoretical work has showen that
the magnon BEC in ferrimagnets and ferromagnets 
remains stable even in the presence of the magnon-magnon interaction
as long as the magnets have a sublattice structure~\cite{NA-Ferri}.
However,
this result is not directly applicable to an antiferromagnet
due to the following essential difference: 
low-energy magnons of an antiferromagnet
have band degeneracy~\cite{Anderson-AF-mag},
whereas low-energy magnons of a ferrimagnet or a ferromagnet
are described by a single band~\cite{Nakamura,HP,NA-Ferri}.
Because of this difference  
the magnon BEC for an antiferromagnet gets a multicomponent order parameter; 
the order parameter for a ferrimagnet or ferromagnet is
one-component~\cite{NA-Ferri,MagBEC-theory}. 
Then this multicomponent BEC 
can be converted into the one-component BEC
by using an external magnetic field that lifts the band degeneracy. 
These properties indicate that
an antiferromagnet provides new opportunities to study
the multicomponent BEC of magnons and the unique magnetic-field effect. 
Nevertheless,
the stability against the interaction remains unclear.

In this paper
we study the BEC of interacting magnons 
in a two-sublattice antiferromagnet and
show a mechanism for controlling its stability. 
Our antiferromagnet is described by
a spin Hamiltonian consisting of
the antiferromagnetic Heisenberg interaction between nearest-neighbor spins,
the single-ion anisotropy,
and the external magnetic field.
After expressing this Hamiltonian in terms of magnon operators
and remarking on several properties,
we formulate an effective theory of the BEC of interacting magnons for the antiferromagnet
and then study the stability in the absence and the presence of the external magnetic field.
We show that
the magnon BEC is unstable
in the absence of the external magnetic field due to
the attractive interband interaction,
which is larger than the repulsive intraband interaction.
We also show that
when the external magnetic field exceeds a critical value,
the repulsive intraband interaction becomes larger than
the attractive interband interaction,
and thus the magnon BEC remains stable.
The main results are summarized in Table \ref{tab}.

We believe that
our theory can study the interaction effects on the stability of 
the BEC of quasiequilibrium magnons in the antiferromagnet.
For realizing the magnon BEC,
it is necessary to adjust the chemical potential of a magnon, $\mu$,
so that $\mu$ satisfies  
$\epsilon_{\textrm{min}}-\mu=0$,
where $\epsilon_{\textrm{min}}$ represents the lowest energy of magnon bands.
This can be realized, for example, using an external pumping~\cite{MagBEC-Nature}.
Although in the presence of the external pumping the system is generally nonequilibrium,
it is possible to realize the quasiequilibrium state
in which the distribution of a magnon can be well approximated by
the Bose distribution function with finite $\mu$; 
in such a quasiequilibrium state
the main effect of the external pumping is to change the value of $\mu$.
However,
even if $\mu$ is adjusted so that
$\epsilon_{\textrm{min}}-\mu=0$ is satisfied for noninteracting magnons,
the magnon BEC becomes unstable for the attractive interaction. 
Since a system at certain times of a quasiequilibrium state
can be approximately described by the magnons whose distribution function 
is the Bose distribution function with finite $\mu$, 
our theory can study how the interaction affects the stability of the magnon BEC
at a snapshot of the quasiequilibrium state. 

\begin{table*}
  \caption{\label{tab}
    Magnon properties of the antiferromagnet
    without and with the external magnetic field.
    Magnon bands are degenerate only for $h=0$.
    The order parameter of the magnon BEC for $h=0$ is two-component,
    whereas that for $h\neq 0$ is one-component.
    The interaction between magnons consists of the intraband interaction
    and the interband interaction, which are repulsive and attractive, respectively.
    If the external magnetic field satisfies $|h| > h_{\textrm{c}}$,
    the repulsive intraband interaction is larger and the magnon BEC remains stable;
    otherwise,
    the attractive interband interaction is larger and the magnon BEC becomes unstable.
  }
  \begin{ruledtabular}
    \begin{tabular}{cccc}
       & $h=0$ & $|h| < h_{\textrm{c}}$ & $|h| > h_{\textrm{c}}$\\ \hline
      Band degeneracy & Degenerate & Nondegenerate & Nondegenerate \\
      Order parameter & Two-component & One-component & One-component\\
      Intraband interaction & Repulsive & Repulsive & Repulsive\\
      Interband interaction & Attractive & Attractive & Attractive\\
      Larger interaction & Interband & Interband & Intraband\\
      Stability of the magnon BEC & Unstable & Unstable & Stable
    \end{tabular}
  \end{ruledtabular}
\end{table*}

\section{Hamiltonian}

Our antiferromagnet is described by the following spin Hamiltonian:
\begin{equation}
  H=2J\sum\limits_{\langle \boldi,\boldj\rangle}\boldS_{\boldi}\cdot\boldS_{\boldj}
  -K\sum\limits_{\boldl}(S_{\boldl}^{z})^{2}
  -h\sum\limits_{\boldl}S_{\boldl}^{z},\label{eq:Hspin}
\end{equation}
where the first term represents the antiferromagnetic Heisenberg interaction
between nearest-neighbor spins,
the second term ($K > 0$) represents
the uniaxial anisotropy energy of a spin for $S > 1/2$, 
and the third term represents the Zeeman energy of the external magnetic field; 
$\sum_{\langle \boldi,\boldj\rangle}$
is the sum for nearest-neighbor spins for $\boldi\in A$ and $\boldj \in B$,
where $A$ and $B$ denote $A$ and $B$ sublattices, 
and $\sum_{\boldl}$ is the sum over all sites; 
and the number of each sublattice is a half of the number of sites $N$. 
The above Hamiltonian is a simple model of a collinear antiferomagnet 
for $\langle \boldS_{\boldi\in A}\rangle={}^{t}(0\ 0\ S)$ and
$\langle \boldS_{\boldj\in B}\rangle={}^{t}(0\ 0\ -S)$~\cite{Yosida}.
Actually,
the antiferromagnetically ordered state of MnF$_{2}$
can be well described by
the spin Hamiltonian consisting of the nearest-neighbor antiferromagnetic
Heisenberg interaction with the small anisotropy~\cite{AF-exp}; 
the estimated value is $K/J \approx 0.1$~\cite{AF-exp}.
We thus believe that 
our model is applicable to MnF$_{2}$ in the presence of the external magnetic field. 
Hereafter we consider cases for $J \gg K, h$.

We have considered not only the Heisenberg interaction
but also the Zeeman energy and the uniaxial anisotropy energy
because the effects of the latter terms are important in antiferromagnets even for $J \gg K, h$.
In the ferrimagnet studied in Ref. \onlinecite{NA-Ferri} 
we did not consider the latter terms 
because their effects are less important than in antiferromagnets.
This difference arises from the following difference in the magnon bands:
in the ferrimagnet, the magnon bands are split even for $h=0$
and the energy splitting is typically of order $J$;
in the antiferromagnet,
the magnon bands are split only for $h\neq 0$
and the energy splitting is of order $h$ [see Eq. (\ref{eq:Hnon-band})].
Since the main effect of the $h$ is to change
the value of the energy splitting of magnon bands,
its effect for the ferrimagnet is negligible compared with that for the antiferromagnet.
Then
the uniaxial anisotropy is necessary to keep the magnon energies for $h\neq 0$ nonnegative. 

We can express 
the above Hamiltonian in terms of
magnon operators by using the Holstein-Primakoff transformation~\cite{HP,Kubo-AF-mag}.
This transformation is expressed as
\begin{align}
  S_{\boldi}^{z}=S-a_{\boldi}^{\dagger}a_{\boldi},
  S_{\boldi}^{-}=a_{\boldi}^{\dagger}\sqrt{2S-a_{\boldi}^{\dagger}a_{\boldi}},
  S_{\boldi}^{+}=(S_{\boldi}^{-})^{\dagger},\label{eq:HP-A}\\
  S_{\boldj}^{z}=-S+b_{\boldj}^{\dagger}b_{\boldj},
  S_{\boldj}^{+}=b_{\boldj}^{\dagger}\sqrt{2S-b_{\boldj}^{\dagger}b_{\boldj}},
  S_{\boldj}^{-}=(S_{\boldj}^{+})^{\dagger},\label{eq:HP-B}
\end{align}
where 
$a_{\boldi}^{\dagger}$ and $a_{\boldi}$
are the creation and annihilation operators of a magnon for $\boldi\in A$,
and $b_{\boldj}^{\dagger}$ and $b_{\boldj}$
are those of a magnon for $\boldj\in B$.
By substituting these equations into Eq. (\ref{eq:Hspin}), 
we obtain the magnon Hamiltonian. 
Since our aim is to clarify the interaction effects
on the magnon BEC, stabilized by the kinetic energy, 
we consider the kinetic energy terms and the dominant terms of the magnon-magnon interaction.
These terms can be derived from the standard calculation~\cite{Oguchi,Nakamura,NA-Ferri}
(for the details see Appendix A).
As a result, 
the kinetic energy terms are given
by the following quadratic terms:
\begin{align}
  H_{\textrm{KE}}=\sum\limits_{\boldq}
  (a_{\boldq}^{\dagger}\ b_{\boldq})
  \left(
  \begin{array}{@{\,}cc@{\,}}
    A(\boldq)+h & C(\boldq) \\
    C(\boldq) & A(\boldq)-h
  \end{array}
  \right)
  \left(
  \begin{array}{@{\,}c@{\,}}
    a_{\boldq} \\
    b_{\boldq}^{\dagger} 
  \end{array}
  \right),\label{eq:Hnon}
\end{align}
where 
$a_{\boldq}=\sqrt{\frac{2}{N}}\sum_{\boldi}e^{-i\boldq\cdot\boldi}a_{\boldi}$,
$b_{\boldq}^{\dagger}=\sqrt{\frac{2}{N}}\sum_{\boldj}e^{-i\boldq\cdot\boldj}b_{\boldj}^{\dagger}$,
$A(\boldq)=2S[J(\boldzero)+K]$,
$C(\boldq)=2SJ(\boldq)$, 
$J(\boldq)=J\sum_{\bolddelta}e^{i\boldq\cdot\bolddelta}$, 
and $\bolddelta$ is a vector to nearest neighbors.
The interaction terms are then given
by the following quartic terms:
\begin{align}
  H_{\textrm{int}}=&-\frac{2}{N}\sum\limits_{\boldq,\boldq^{\prime}}
  [J(\boldzero)a_{\boldq}^{\dagger}a_{\boldq}b_{\boldq^{\prime}}^{\dagger}b_{\boldq^{\prime}}
    +J(\boldq-\boldq^{\prime})a_{\boldq}^{\dagger}a_{\boldq^{\prime}}b_{\boldq}^{\dagger}b_{\boldq^{\prime}}\notag\\
    +&J(\boldq)a_{\boldq}b_{\boldq^{\prime}}^{\dagger}b_{\boldq}b_{\boldq^{\prime}}
    +J(\boldq)a_{\boldq^{\prime}}^{\dagger}a_{\boldq}a_{\boldq^{\prime}}b_{\boldq}]
  +(\textrm{H.c.}).\label{eq:Hint}
\end{align}
The interaction terms consist of the quartic terms arising from the Heisenberg interaction
because we have considered the interaction between magnons at different sites;
in Eq. (\ref{eq:Hspin}) only the first term gives such an interaction.

\section{Properties of magnons}

The essential difference between the antiferromagnet
and a ferromagnet or a ferrimagnet
is degeneracy of magnon bands. 
To see this, we diagonalize Eq. (\ref{eq:Hnon}).
This diagonalization can be performed using the Bogoliubov transformation,
\begin{equation}
  \left(
  \begin{array}{@{\,}c@{\,}}
    a_{\boldq} \\
    b_{\boldq}^{\dagger} 
  \end{array}
  \right)
  =
  \left(
  \begin{array}{@{\,}cc@{\,}}
    c_{\boldq}  & -s_{\boldq} \\
    -s_{\boldq} & c_{\boldq} 
  \end{array}
  \right)
  \left(
  \begin{array}{@{\,}c@{\,}}
    \alpha_{\boldq} \\
    \beta_{\boldq}^{\dagger} 
  \end{array}
  \right),\label{eq:Bogoliubov-trans}
\end{equation}
where $c_{\boldq}=\cosh\theta_{\boldq}$ and $s_{\boldq}=\sinh\theta_{\boldq}$.
Actually, by substituting Eq. (\ref{eq:Bogoliubov-trans}) into Eq. (\ref{eq:Hnon}) 
and setting $\tanh 2\theta_{\boldq}=C(\boldq)/A(\boldq)$, 
we obtain
\begin{align}
  H_{\textrm{KE}}=\sum\limits_{\boldq}
    (\alpha_{\boldq}^{\dagger}\ \beta_{\boldq})
  \left(
  \begin{array}{@{\,}cc@{\,}}
    \epsilon(\boldq)+h & 0 \\
    0 & \epsilon(\boldq)-h
  \end{array}
  \right)
  \left(
  \begin{array}{@{\,}c@{\,}}
    \alpha_{\boldq} \\
    \beta_{\boldq}^{\dagger} 
  \end{array}
  \right),\label{eq:Hnon-band}
\end{align}
where $\epsilon(\boldq)=\sqrt{A(\boldq)^{2}-C(\boldq)^{2}}$.
Equation (\ref{eq:Hnon-band}) shows that
the $\alpha$ and $\beta$ bands are degenerate for $h=0$.
This property is distinct from 
a lack of such degeneracy in ferromagnets and ferrimagnets;
for example,
in a ferrimagnet with a similar sublattice structure
two bands are nondegenerate even for $h=0$~\cite{Nakamura,NA-Ferri}. 
This difference results from the difference in time-reversal symmetry,
which holds only for the antiferromagnet for $h=0$. 
Actually,
the external magnetic field, which breaks time-reversal symmetry, 
lifts the band degeneracy, as seen from Eq. (\ref{eq:Hnon-band}). 

The above properties result in
two unique properties of the magnon BEC in the antiferromagnet.  
Since $\epsilon(\boldq)=2S\sqrt{[J(\boldzero)+K]^{2}-J(\boldq)^{2}}$,
$\epsilon(\boldzero)$ is the lowest energy.
A combination of this and the band degeneracy for $h=0$ 
indicates that
in the magnon BEC for $h=0$
a macroscopic number of magnons 
occupies the $\boldq=\boldzero$ states of the $\alpha$ and $\beta$ bands.
Therefore
the magnon BEC in the antiferromagnet for $h=0$
possesses a multicomponent order parameter
(i.e., $\langle \alpha_{\boldzero}\rangle$ and $\langle \beta_{\boldzero}^{\dagger}\rangle$).
This is in contrast with
the magnon BEC in a ferromagnet and a ferrimagnet
because that is described by a one-component order parameter~\cite{NA-Ferri,MagBEC-theory}. 
Then, 
by using the external magnetic field,
the magnon BEC in the antiferromagnet
can change from multicomponent BEC 
to one-component BEC 
as a result of lifting the band degeneracy.
For example,
for $h < 0$
the lowest-energy state is the $\boldq=\boldzero$ state of the $\alpha$ band 
and the order parameter is $\langle \alpha_{\boldzero}\rangle$.

\section{Interaction effects on the stability of the magnon BEC}

We now formulate the effective theory of 
the BEC of interacting magnons in the antiferromagnet for $h=0$.
Since we can express the terms in Eq. (\ref{eq:Hint})
in terms of the operators of the $\alpha$ and $\beta$ bands
by using Eq. (\ref{eq:Bogoliubov-trans}),
we can write $H_{\textrm{int}}$ as the sum of intraband terms and interband terms, 
$H_{\textrm{int}}=H_{\textrm{intra}}+H_{\textrm{inter}}$, 
where
$H_{\textrm{intra}}$ consists of the intraband terms of the $\alpha$ band and of the $\beta$ band  
and
$H_{\textrm{inter}}$ consists of the interband terms between these bands.
The expressions for $H_{\textrm{intra}}$ and $H_{\textrm{inter}}$
are derived in Appendix B. 
Then,
since the main effects of the interaction terms can be taken into account
in the mean-field approximation~\cite{Nakamura,Oguchi,NA-Ferri},
we can approximate the terms of $H_{\textrm{intra}}$ and $H_{\textrm{inter}}$
by the mean-field interaction terms. 
Using these procedures and performing a calculation
similar to that for the ferrimagnet~\cite{NA-Ferri},  
we can obtain the leading terms of $H_{\textrm{intra}}$ and $H_{\textrm{inter}}$.
First,
as we show in Appendix C,
the contributions from 
the mean-field interaction terms for $\boldq=\boldq^{\prime}=\boldzero$,
for $\boldq=\boldzero$, $\boldq^{\prime}\neq\boldzero$,
and for $\boldq\neq\boldzero$, $\boldq^{\prime}=\boldzero$ 
are zero within the leading order. 
Then, as we derive in Appendix D,  
the mean-field interaction terms for $\boldq\neq\boldzero$, $\boldq^{\prime}\neq\boldzero$
are given as follows:
\begin{widetext}
\begin{align}
  &H_{\textrm{intra}}=
  \frac{1}{N}
  \sum\limits_{\boldq,\boldq^{\prime}\neq \boldzero}
  \Gamma_{\textrm{intra1}}(\boldq,\boldq^{\prime})
  (n_{\boldq^{\prime}\alpha}\alpha_{\boldq}^{\dagger}\alpha_{\boldq}
  +n_{\boldq^{\prime}\beta}\beta_{\boldq}^{\dagger}\beta_{\boldq}),\label{eq:Hintra}\\
  &H_{\textrm{inter}}=
  \frac{1}{N}
  \sum\limits_{\boldq,\boldq^{\prime}\neq \boldzero}
  \Gamma_{\textrm{inter}1}(\boldq,\boldq^{\prime})
  (n_{\boldq^{\prime}\beta}\alpha_{\boldq}^{\dagger}\alpha_{\boldq}
  +n_{\boldq^{\prime}\alpha}\beta_{\boldq}^{\dagger}\beta_{\boldq})
  +\frac{1}{N}
  \sum\limits_{\boldq\neq \boldzero}
  \Gamma_{\textrm{inter}2}(\boldq)
  (n_{\boldq\beta}\alpha_{\boldq}^{\dagger}\alpha_{\boldq}
  +n_{\boldq\alpha}\beta_{\boldq}^{\dagger}\beta_{\boldq})\notag\\
  &+\frac{1}{N}
  \sum\limits_{\boldq,\boldq^{\prime}\neq \boldzero}
  \Gamma_{\textrm{inter}3}(\boldq,\boldq^{\prime})
  (n_{\boldq^{\prime}\alpha}+n_{\boldq^{\prime}\beta})
  (\alpha_{\boldq}\beta_{\boldq}+\alpha_{\boldq}^{\dagger}\beta_{\boldq}^{\dagger})
  +\frac{1}{N}
  \sum\limits_{\boldq\neq \boldzero}
  \Gamma_{\textrm{inter}4}(\boldq)
  (n_{\boldq\alpha}+n_{\boldq\beta})
  (\alpha_{\boldq}\beta_{\boldq}+\alpha_{\boldq}^{\dagger}\beta_{\boldq}^{\dagger}),\label{eq:Hinter}
\end{align} 
where
\begin{align}
  &\Gamma_{\textrm{intra}1}(\boldq,\boldq^{\prime})
  =-4J(\boldzero)(c_{\boldq}^{2}s_{\boldq^{\prime}}^{2}+c_{\boldq^{\prime}}^{2}s_{\boldq}^{2})
  -8J(\boldq-\boldq^{\prime})c_{\boldq}s_{\boldq}c_{\boldq^{\prime}}s_{\boldq^{\prime}}
  +4J(\boldq)c_{\boldq}s_{\boldq}(c_{\boldq^{\prime}}^{2}+s_{\boldq^{\prime}}^{2})
  +4J(\boldq^{\prime})c_{\boldq^{\prime}}s_{\boldq^{\prime}}(c_{\boldq}^{2}+s_{\boldq}^{2}),\label{eq:Gam-intra1}\\
  &\Gamma_{\textrm{inter}1}(\boldq,\boldq^{\prime})
  =-4J(\boldzero)(c_{\boldq}^{2}c_{\boldq^{\prime}}^{2}+s_{\boldq^{\prime}}^{2}s_{\boldq}^{2})
  -8J(\boldq-\boldq^{\prime})c_{\boldq}s_{\boldq}c_{\boldq^{\prime}}s_{\boldq^{\prime}}
  +4J(\boldq)c_{\boldq}s_{\boldq}(c_{\boldq^{\prime}}^{2}+s_{\boldq^{\prime}}^{2})
  +4J(\boldq^{\prime})c_{\boldq^{\prime}}s_{\boldq^{\prime}}(c_{\boldq}^{2}+s_{\boldq}^{2}),\label{eq:Gam-inter1}\\
  &\Gamma_{\textrm{inter}2}(\boldq)
  =-4J(\boldzero)(c_{\boldq}^{2}+s_{\boldq}^{2})^{2}
  +8J(\boldq)c_{\boldq}s_{\boldq}(c_{\boldq}^{2}+s_{\boldq}^{2}),\label{eq:Gam-inter2}\\
  &\Gamma_{\textrm{inter}3}(\boldq,\boldq^{\prime})
  =4J(\boldzero)c_{\boldq}s_{\boldq}(c_{\boldq^{\prime}}^{2}+s_{\boldq^{\prime}}^{2})
  +4J(\boldq-\boldq^{\prime})c_{\boldq^{\prime}}s_{\boldq^{\prime}}(c_{\boldq}^{2}+s_{\boldq}^{2})
  -2J(\boldq)(c_{\boldq}^{2}+s_{\boldq}^{2})(c_{\boldq^{\prime}}^{2}+s_{\boldq^{\prime}}^{2})
  -8J(\boldq^{\prime})c_{\boldq}s_{\boldq}c_{\boldq^{\prime}}s_{\boldq^{\prime}},\label{eq:Gam-inter3}\\
  &\Gamma_{\textrm{inter}4}(\boldq)
  =8J(\boldzero)c_{\boldq}s_{\boldq}(c_{\boldq}^{2}+s_{\boldq}^{2})
  -2J(\boldq)(c_{\boldq}^{2}+s_{\boldq}^{2})^{2}
  -8J(\boldq)c_{\boldq}^{2}s_{\boldq}^{2}.\label{eq:Gam-inter4}
\end{align}
\end{widetext}
In Eqs. (\ref{eq:Hintra}) and (\ref{eq:Hinter})
we have defined
$n_{\boldq\alpha}=\langle\alpha_{\boldq}^{\dagger}\alpha_{\boldq}\rangle=n[\epsilon(\boldq)]$ and 
$n_{\boldq\beta}=\langle\beta_{\boldq}^{\dagger}\beta_{\boldq}\rangle=n[\epsilon(\boldq)]$,
where $n(\epsilon)$ is the Bose distribution function.
Then, since dominant low-energy excitations are described by 
low-$\boldq$ magnons of the $\alpha$ and $\beta$ bands,
we can estimate 
$\Gamma_{\textrm{intra}1}(\boldq,\boldq^{\prime})$,
$\Gamma_{\textrm{inter}1}(\boldq,\boldq^{\prime})$,
$\Gamma_{\textrm{inter}2}(\boldq)$,
$\Gamma_{\textrm{inter}3}(\boldq,\boldq^{\prime})$,
and $\Gamma_{\textrm{inter}4}(\boldq)$
in the limit $|\boldq|,|\boldq^{\prime}|\rightarrow 0$. 
After the calculation described in Appendix E, 
we obtain
\begin{align}
&\Gamma_{\textrm{intra}1}(\boldq,\boldq^{\prime})\sim 2J(\boldzero),\label{eq:Gm-intra1}\\ 
&\Gamma_{\textrm{inter}1}(\boldq,\boldq^{\prime})\sim -2J(\boldzero),\label{eq:Gm-inter1}\\
&\Gamma_{\textrm{inter}2}(\boldq)\sim -2J(\boldzero),\label{eq:Gm-inter2}\\
&\Gamma_{\textrm{inter}3}(\boldq,\boldq^{\prime})\sim 0,\label{eq:Gm-inter3}\\
&\Gamma_{\textrm{inter}4}(\boldq)\sim 0.\label{eq:Gm-inter4}
\end{align}
Thus 
the leading term of the intraband interaction is repulsive,
whereas that of the interband interaction is attractive.
In addition, 
Eqs. (\ref{eq:Gm-intra1}){--}(\ref{eq:Gm-inter4}) 
with Eqs. (\ref{eq:Hintra}) and (\ref{eq:Hinter}) 
show that
the leading terms of
$H_{\textrm{intra}}$ and $H_{\textrm{inter}}$
are written as follows:
\begin{align}
  &H_{\textrm{intra}}=
  \frac{2}{N}J(\boldzero)
  \sum\limits_{\boldq,\boldq^{\prime}\neq \boldzero}
  (n_{\boldq^{\prime}\alpha}\alpha_{\boldq}^{\dagger}\alpha_{\boldq}
  +n_{\boldq^{\prime}\beta}\beta_{\boldq}^{\dagger}\beta_{\boldq}),\label{eq:Hintra-last}\\
  &H_{\textrm{inter}}=
  -\frac{2}{N}J(\boldzero)
  \sum\limits_{\boldq,\boldq^{\prime}\neq \boldzero}
  (n_{\boldq^{\prime}\beta}\alpha_{\boldq}^{\dagger}\alpha_{\boldq}
  +n_{\boldq^{\prime}\alpha}\beta_{\boldq}^{\dagger}\beta_{\boldq})\notag\\
  &-\frac{2}{N}J(\boldzero)
  \sum\limits_{\boldq\neq \boldzero}
  (n_{\boldq\beta}\alpha_{\boldq}^{\dagger}\alpha_{\boldq}
  +n_{\boldq\alpha}\beta_{\boldq}^{\dagger}\beta_{\boldq}).\label{eq:Hinter-last}
\end{align} 
Combining Eqs. (\ref{eq:Hintra-last}) and (\ref{eq:Hinter-last})
with Eq. (\ref{eq:Hnon-band}) for $h=0$ 
and using $n_{\boldq\alpha}=n_{\boldq\beta}=n[\epsilon(\boldq)]$, 
we obtain the following effective Hamiltonian: 
\begin{align}
  H_{\textrm{eff}}=
  \sum\limits_{\boldq}
  \epsilon^{\ast}(\boldq)(\alpha_{\boldq}^{\dagger}\alpha_{\boldq}
  +\beta_{\boldq}^{\dagger}\beta_{\boldq}),
\end{align}
where
\begin{align}
  \epsilon^{\ast}(\boldq)=\epsilon(\boldq)
  +(1-\delta_{\boldq,\boldzero})\frac{1}{N}\sum_{\boldq^{\prime}\neq \boldzero}
  \Gamma_{\textrm{eff}}(\boldq,\boldq^{\prime})n[\epsilon(\boldq^{\prime})],\label{eq:e^ast}
\end{align}
and
\begin{align}
  \Gamma_{\textrm{eff}}(\boldq,\boldq^{\prime})
  =&\Gamma_{\textrm{intra1}}(\boldq,\boldq^{\prime})
  +\Gamma_{\textrm{inter}1}(\boldq,\boldq^{\prime})
  +\delta_{\boldq^{\prime},\boldq}\Gamma_{\textrm{inter}2}(\boldq)\notag\\
  \sim & -2J(\boldzero)\delta_{\boldq^{\prime},\boldq}.\label{eq:Gm^eff}
\end{align}

By using the above effective theory, 
we can show that
the attractive interband interaction causes
the magnon BEC for $h=0$ to be unstable.
If $\epsilon^{\ast}(\boldzero)=\epsilon(\boldzero)$
remains the lowest-energy state,
the magnon BEC remains stable even in the presence of the magnon-magnon interaction;
otherwise, the magnon BEC becomes unstable.
The former condition is satisfied
if $\Gamma_{\textrm{eff}}(\boldq,\boldq^{\prime})$ is positive,
i.e., the effective interaction is repulsive.
Note that $n[\epsilon(\boldq^{\prime})]$ is non-negative. 
From Eq. (\ref{eq:e^ast}) with Eq. (\ref{eq:Gm^eff}),
we can deduce that 
the effective interaction becomes attractive 
and it makes the magnon BEC unstable. 

In a similar way 
we analyze the interaction effects on the magnon BEC for $h\neq 0$
and show the condition for the stability. 
We consider the case for $h < 0$ 
because the case for $h > 0$ can be similarly analyzed.
For $h < 0$
the band degeneracy is lifted
and the low-energy magnons are described by the $\alpha$ band.
Since the effect of $h$ is to replace
the energy dispersions of noninteracting magnons of the $\alpha$ and $\beta$ bands
by $\epsilon_{\alpha}(\boldq)=\epsilon(\boldq)+h$ and $\epsilon_{\beta}(\boldq)=\epsilon(\boldq)-h$,
respectively,
we obtain the effective theory for $h\neq 0$
by replacing $\epsilon(\boldq)$'s for the $\alpha$ and $\beta$ bands
in the effective theory for $h=0$ 
by $\epsilon_{\alpha}(\boldq)$ and $\epsilon_{\beta}(\boldq)$,
respectively;
the expressions of $\Gamma_{\textrm{intra}1}(\boldq,\boldq^{\prime})$,
$\Gamma_{\textrm{inter}1}(\boldq,\boldq^{\prime})$,
$\Gamma_{\textrm{inter}2}(\boldq)$, $\Gamma_{\textrm{inter}3}(\boldq,\boldq^{\prime})$,
and $\Gamma_{\textrm{inter}4}(\boldq)$ remain unchanged
because $c_{\boldq}^{2}$, $s_{\boldq}^{2}$, and $c_{\boldq}s_{\boldq}$ are
independent of $h$
[see Eqs. (\ref{eq:iden1}){--}(\ref{eq:sinh2})]. 
Therefore
the effective theory for $h<0$ is described by the following Hamiltonian:
\begin{align}
  H_{\textrm{eff}}=
  \sum\limits_{\boldq}
  \epsilon_{\alpha}^{\ast}(\boldq)\alpha_{\boldq}^{\dagger}\alpha_{\boldq},
\end{align}
where
\begin{align}
  &\epsilon_{\alpha}^{\ast}(\boldq)=
  \epsilon_{\alpha}(\boldq)
  +(1-\delta_{\boldq,\boldzero})\frac{2}{N}J(\boldzero)\sum_{\boldq^{\prime}\neq \boldzero}\notag\\
  &\times \Bigl\{n[\epsilon_{\alpha}(\boldq^{\prime})]
  -(1+\delta_{\boldq^{\prime},\boldq})n[\epsilon_{\beta}(\boldq^{\prime})]
  \Bigr\}.\label{eq:del-epsilon}
\end{align}
The above effective Hamiltonian includes not only the intraband interaction of the $\alpha$ band
but also the interband interaction between the $\alpha$ and $\beta$ bands 
because the energy difference between these bands is of order $h$, much smaller than $J$.
Equation (\ref{eq:del-epsilon}) shows that
the magnon BEC is unstable for
$n[\epsilon_{\alpha}(\boldq^{\prime})] < (1+\delta_{\boldq^{\prime},\boldq})n[\epsilon_{\beta}(\boldq^{\prime})]$,
which means that the repulsive intraband interaction is smaller
than the attractive interband interaction, 
whereas the magnon BEC is stable for  
$n[\epsilon_{\alpha}(\boldq^{\prime})] > (1+\delta_{\boldq^{\prime},\boldq})n[\epsilon_{\beta}(\boldq^{\prime})]$,
which means that the repulsive intraband interaction is larger.
Since $n(\epsilon)$ is a monotonically decreasing function
and an increase of $|h|$ leads to both 
an increase of $n[\epsilon_{\alpha}(\boldq^{\prime})]=n[\epsilon(\boldq^{\prime})-|h|]$
and a decrease of $n[\epsilon_{\beta}(\boldq^{\prime})]=n[\epsilon(\boldq^{\prime})+|h|]$
at each $\epsilon(\boldq^{\prime})$,
there should be a critical value of $|h|$, $h_{\textrm{c}}$, which is defined as
$n[\epsilon(\boldq^{\prime})-h_{\textrm{c}}]
= (1+\delta_{\boldq^{\prime},\boldq})n[\epsilon(\boldq^{\prime})+h_{\textrm{c}}]$;
this can be seen also from
estimating the terms in the curly brackets of Eq. (\ref{eq:del-epsilon})
to the leading term in $h$~\cite{remark}. 
Therefore
the BEC of interacting magnons remains stable for $|h| > h_{\textrm{c}}$,
whereas it is unstable for $|h| < h_{\textrm{c}}$.

\section{Discussion}

To get a deeper understanding,
we compare our results with the results
in ferrimagnets and ferromagnets~\cite{NA-Ferri,MagBEC-theory}.
First, 
our result for $h=0$ is distinct from the results
in ferrimagnets and ferromagnets~\cite{NA-Ferri,MagBEC-theory}
because multicomponent BEC is characteristic of the antiferromagnet. 
We thus believe that
our result is the first step towards understanding
the interaction effects in multicompoment BEC of magnons. 
Second,
our result for $|h| > h_{\textrm{c}}$ is similar to 
the results in the two-sublattice ferrimagnet and ferromagnet~\cite{NA-Ferri}
because these show that 
the one-component BEC of interacting magnons remains stable
due to the repulsive intraband interaction.
The origin of this repulsive interaction can be understood in the same manner:
the magnons in the different sublattices give the different contributions to
the intraband interaction due to the different coefficients
in the Bogoliubov transformation Eq. (\ref{eq:Bogoliubov-trans}).
This similarity suggests that
the magnon BEC may be experimentally observed
for an antiferromagnet with an external magnetic field; 
this suggestion is consistent with experiments~\cite{MagBEC-AF1,MagBEC-AF2}.
Third, 
our result for $|h| < h_{\textrm{c}}$
contrasts with the results in the two-sublattice ferrimagnet and ferromagnet~\cite{NA-Ferri} 
because the former shows that
the one-component BEC of interacting magnons is unstable
despite the existence of the sublattice structure.
This difference arises from
the competition between the repulsive intraband interaction
and the attractive interband interaction in the antiferromagnet.
Note that 
the interband interaction is non-negligible
only if the energy difference between bands is small 
because $n(\epsilon)$ is large only for small $\epsilon$. 
Since that energy difference is typically of order $J$ 
for ferrimagnets and ferromagnets~\cite{NA-Ferri,Nakamura}, 
the interband interaction is negligible for them.
Thus
the competition between the intraband and interband interactions
is characteristic of the antiferromagnet.
From the above comparison
we conclude that 
the interaction effects in the antiferromagnet are more complicated
due to
the band degeneracy for no external magnetic field and 
the small energy difference between magnon bands.

We now discuss the implications of our results. 
Our results show that the stability of the BEC of interacting magnons
in the antiferromagnet is controllable by tuning the external magnetic field.
Therefore 
our mechanism can be used for further studies of not only the stable BEC of interacting magnons 
but also the unstable BEC.
In particular,
our result for $h=0$ provides a unique opportunity for studying
the unstable multicomponent BEC of interacting magnons.
Since our model is applicable to MnF$_{2}$, as described in Sec. II,
these properties could be observed in the magnon BEC for MnF$_{2}$. 
Our results also show that
the sign of the effective interaction between magnons of the antiferromagnet is 
sensitive to the external magnetic field.
This is distinct from the expectation that
the properties of antiferromagnets are insensitive to magnetic fields~\cite{AF-Mag,Review1,Review2}.
Therefore
this distinction suggests that
it is not always correct to neglect the magnetic-field dependence in antiferromagnets.
In particular,
in cases of the magnon properties
that depend on the sign of the effective interaction between magnons, 
it is crucial to study the magnetic-field dependence. 
Then our theory can be extended to more complicated magnets. 
Our study is thus useful for further studies
of the one-component and the multicomponent BEC of interacting magnons in various magnets. 

\section{Summary}

In summary we studied 
the BEC of interacting magnons
in the two-sublattice antiferromagnet without and with the external magnetic field
and revealed a method to control its stability. 
We showed that 
the magnon BEC for no external magnetic field gets the two-component order parameter
due to the band degeneracy 
and this two-component BEC is destabilized by 
the attractive interband interaction, which is larger than
the repulsive intraband interaction.
We also showed that 
the two-component BEC is converted into one-component BEC
by the external magnetic field, which lifts the band degeneracy, 
and this one-component BEC is stabilized 
only if the external magnetic field exceeds the critical value, above which
the repulsive intraband interaction is larger than the attractive interband interaction.


\appendix
\begin{widetext}
\section{Derivation of Eqs. (\ref{eq:Hnon}) and (\ref{eq:Hint})}
We derive Eqs. (\ref{eq:Hnon}) and (\ref{eq:Hint}).
By substituting Eqs. (\ref{eq:HP-A}) and (\ref{eq:HP-B}) into Eq. (\ref{eq:Hspin}), 
we can express the Hamiltonian of Eq. (\ref{eq:Hspin}) in terms of the magnon operators.
Since the quadratic terms correspond to the kinetic energy terms
and the quartic terms correspond to the interaction terms~\cite{Oguchi,Nakamura,NA-Ferri},
we explain the details of the derivations of the quadratic terms and the quartic terms.
First, 
the quadratic terms are given by 
\begin{align}
  H_{\textrm{KE}}
  =2JS\sum\limits_{\langle \boldi,\boldj\rangle}(a_{\boldi}^{\dagger}a_{\boldi}+b_{\boldj}^{\dagger}b_{\boldj}
  +a_{\boldi}b_{\boldj}+a_{\boldi}^{\dagger}b_{\boldj}^{\dagger})
  +(2KS+h)\sum\limits_{\boldi}a_{\boldi}^{\dagger}a_{\boldi}
  +(2KS-h)\sum\limits_{\boldj}b_{\boldj}^{\dagger}b_{\boldj}.\label{eq:HKE}
\end{align}
We can rewrite 
the above quadratic terms by using
$a_{\boldi}=\sqrt{\frac{2}{N}}\sum_{\boldq}e^{i\boldq\cdot\boldi}a_{\boldq}$,
$b_{\boldj}^{\dagger}=\sqrt{\frac{2}{N}}\sum_{\boldq}e^{i\boldq\cdot\boldj}b_{\boldq}^{\dagger}$, and 
$J(\boldq)=J\sum_{\bolddelta}e^{i\boldq\cdot\bolddelta}$.
As a result, we obtain
\begin{align}
  H_{\textrm{KE}}
  =\{2S[J(\boldzero)+K]+h\}\sum\limits_{\boldq}a_{\boldq}^{\dagger}a_{\boldq}
  +\{2S[J(\boldzero)+K]-h\}\sum\limits_{\boldq}b_{\boldq}^{\dagger}b_{\boldq}
  +2S\sum\limits_{\boldq}J(\boldq)(a_{\boldq}b_{\boldq}+a_{\boldq}^{\dagger}b_{\boldq}^{\dagger}).
\end{align}
This is Eq. (\ref{eq:Hnon}).
Then
the quartic terms which arise from the Heisenberg interaction are given by 
\begin{equation}
  H_{\textrm{int}}
  =-2J\sum\limits_{\langle \boldi,\boldj\rangle}a_{\boldi}^{\dagger}a_{\boldi}b_{\boldj}^{\dagger}b_{\boldj}
  -\frac{J}{2}\sum\limits_{\langle \boldi,\boldj\rangle}
  (a_{\boldi}b_{\boldj}^{\dagger}b_{\boldj}b_{\boldj}
  +a_{\boldi}^{\dagger}a_{\boldi}a_{\boldi}b_{\boldj}
  +a_{\boldi}^{\dagger}b_{\boldj}^{\dagger}b_{\boldj}^{\dagger}b_{\boldj}
  +a_{\boldi}^{\dagger}a_{\boldi}^{\dagger}a_{\boldi}b_{\boldj}^{\dagger}).
\end{equation}
The above terms can be divided into two parts:
$H_{\textrm{int}}=V+V^{\dagger}$, where
\begin{equation}
  V=-J\sum\limits_{\langle \boldi,\boldj\rangle}a_{\boldi}^{\dagger}a_{\boldi}b_{\boldj}^{\dagger}b_{\boldj}
  -\frac{J}{2}\sum\limits_{\langle \boldi,\boldj\rangle}
  (a_{\boldi}b_{\boldj}^{\dagger}b_{\boldj}b_{\boldj}
  +a_{\boldi}^{\dagger}a_{\boldi}a_{\boldi}b_{\boldj}).\label{eq:V}
\end{equation}
By using $a_{\boldi}=\sqrt{\frac{2}{N}}\sum_{\boldq}e^{i\boldq\cdot\boldi}a_{\boldq}$,
$b_{\boldj}^{\dagger}=\sqrt{\frac{2}{N}}\sum_{\boldq}e^{i\boldq\cdot\boldj}b_{\boldq}^{\dagger}$, and 
$J(\boldq)=J\sum_{\bolddelta}e^{i\boldq\cdot\bolddelta}$,
we can write Eq. (\ref{eq:V}) as
\begin{equation}
  V=
  -\frac{1}{N}\sum\limits_{\boldq_{1},\boldq_{2},\boldq_{3},\boldq_{4}}
  [2J(\boldq_{1}-\boldq_{3})a_{\boldq_{1}}^{\dagger}a_{\boldq_{3}}b_{\boldq_{4}}^{\dagger}b_{\boldq_{2}}
  +J(\boldq_{1})
  (a_{\boldq_{1}}b_{\boldq_{2}}^{\dagger}b_{\boldq_{3}}b_{\boldq_{4}}
  +a_{\boldq_{2}}^{\dagger}a_{\boldq_{3}}a_{\boldq_{4}}b_{\boldq_{1}})]
  \delta_{\boldq_{1}+\boldq_{2},\boldq_{3}+\boldq_{4}}.
\end{equation}
Since the dominant terms of $V$ come from
the diagonal terms~\cite{Oguchi,Nakamura,NA-Ferri},
the terms
for $\boldq_{1}=\boldq_{3}=\boldq$, $\boldq_{2}=\boldq_{4}=\boldq^{\prime}$
and for $\boldq_{1}=\boldq_{4}=\boldq$,
$\boldq_{2}=\boldq_{3}=\boldq^{\prime}$,
the dominant terms of $V$ are given by
\begin{align}
  V= -\frac{2}{N}\sum\limits_{\boldq,\boldq^{\prime}}
  [J(\boldzero)a_{\boldq}^{\dagger}a_{\boldq}b_{\boldq^{\prime}}^{\dagger}b_{\boldq^{\prime}}
    +J(\boldq-\boldq^{\prime})a_{\boldq}^{\dagger}a_{\boldq^{\prime}}b_{\boldq}^{\dagger}b_{\boldq^{\prime}}
    +J(\boldq)a_{\boldq}b_{\boldq^{\prime}}^{\dagger}b_{\boldq}b_{\boldq^{\prime}}
    +J(\boldq)a_{\boldq^{\prime}}^{\dagger}a_{\boldq}a_{\boldq^{\prime}}b_{\boldq}].
\end{align}
Combining this equation with $H_{\textrm{int}}=V+V^{\dagger}$, we obtain
Eq. (\ref{eq:Hint}).

\section{$H_{\textrm{intra}}$ and $H_{\textrm{inter}}$ expressed
in terms of the operators of the $\alpha$- and $\beta$-band magnons}

We rewrite $H_{\textrm{int}}$ as the sum of $H_{\textrm{intra}}$ and $H_{\textrm{inter}}$
by using Eq. (\ref{eq:Bogoliubov-trans}). 
Since $H_{\textrm{int}}$ consists of
the terms quartic in the operators of $A$- and $B$-sublattice magnons,
the terms of $H_{\textrm{int}}$ expressed in terms of the operators
of the $\alpha$- and $\beta$-band magnons can be divided into five parts:
products of four operators of the $\alpha$-band magnons,
products of four operators of the $\beta$-band magnons,
products of two operators of the $\alpha$-band magnons
and two operators of the $\beta$-band magnons,
products of three operators of the $\alpha$-band magnons
and one operator of the $\beta$-band magnons,
and products of one operator of the $\alpha$-band magnons
and three operators of the $\beta$-band magnons.
The first two parts, $H_{\alpha\alpha\alpha\alpha}$ and $H_{\beta\beta\beta\beta}$,
lead to the intraband interaction, $H_{\textrm{intra}}=H_{\alpha\alpha\alpha\alpha}+H_{\beta\beta\beta\beta}$, 
whereas the latter three parts,
$H_{\alpha\alpha\beta\beta}$, $H_{\alpha\alpha\alpha\beta}$, and $H_{\alpha\beta\beta\beta}$,
lead to the interband interaction,
$H_{\textrm{inter}}=H_{\alpha\alpha\beta\beta}+H_{\alpha\alpha\alpha\beta}+H_{\alpha\beta\beta\beta}$.
By expressing $H_{\textrm{int}}$ in terms of the operators of
the $\alpha$- and $\beta$-band magnons,
we obtain $H_{\alpha\alpha\alpha\alpha}=V_{1}+V_{1}^{\dagger}$, $H_{\beta\beta\beta\beta}=V_{2}+V_{2}^{\dagger}$,
$H_{\alpha\alpha\beta\beta}=V_{3}+V_{3}^{\dagger}$, $H_{\alpha\alpha\alpha\beta}=V_{4}+V_{4}^{\dagger}$,
and $H_{\alpha\beta\beta\beta}=V_{5}+V_{5}^{\dagger}$, where 
\begin{align}
  V_{1}=&
  -\frac{2}{N}\sum\limits_{\boldq,\boldq^{\prime}}
  J(\boldzero)c_{\boldq}^{2}s_{\boldq^{\prime}}^{2}
  \alpha_{\boldq}^{\dagger}\alpha_{\boldq}\alpha_{\boldq^{\prime}}\alpha_{\boldq^{\prime}}^{\dagger}
  -\frac{2}{N}\sum\limits_{\boldq,\boldq^{\prime}}
  J(\boldq-\boldq^{\prime})c_{\boldq}s_{\boldq}c_{\boldq^{\prime}}s_{\boldq^{\prime}}
  \alpha_{\boldq}^{\dagger}\alpha_{\boldq^{\prime}}\alpha_{\boldq}\alpha_{\boldq^{\prime}}^{\dagger}
  +\frac{2}{N}\sum\limits_{\boldq,\boldq^{\prime}}J(\boldq)c_{\boldq}s_{\boldq}s_{\boldq^{\prime}}^{2}
  \alpha_{\boldq}\alpha_{\boldq^{\prime}}\alpha_{\boldq}^{\dagger}\alpha_{\boldq^{\prime}}^{\dagger}\notag\\
  &+\frac{2}{N}\sum\limits_{\boldq,\boldq^{\prime}}J(\boldq)c_{\boldq}s_{\boldq}c_{\boldq^{\prime}}^{2}
  \alpha_{\boldq^{\prime}}^{\dagger}\alpha_{\boldq}\alpha_{\boldq^{\prime}}\alpha_{\boldq}^{\dagger},\label{eq:V1}\\
  V_{2}=&
  -\frac{2}{N}\sum\limits_{\boldq,\boldq^{\prime}}
  J(\boldzero)c_{\boldq^{\prime}}^{2}s_{\boldq}^{2}
  \beta_{\boldq}\beta_{\boldq}^{\dagger}\beta_{\boldq^{\prime}}^{\dagger}\beta_{\boldq^{\prime}}
  -\frac{2}{N}\sum\limits_{\boldq,\boldq^{\prime}}  
  J(\boldq-\boldq^{\prime})c_{\boldq}s_{\boldq}c_{\boldq^{\prime}}s_{\boldq^{\prime}}
  \beta_{\boldq}\beta_{\boldq^{\prime}}^{\dagger}\beta_{\boldq}^{\dagger}\beta_{\boldq^{\prime}}
  +\frac{2}{N}\sum\limits_{\boldq,\boldq^{\prime}}
  J(\boldq)c_{\boldq}s_{\boldq}c_{\boldq^{\prime}}^{2}
  \beta_{\boldq}^{\dagger}\beta_{\boldq^{\prime}}^{\dagger}\beta_{\boldq}\beta_{\boldq^{\prime}}
  \notag\\
  &+\frac{2}{N}\sum\limits_{\boldq,\boldq^{\prime}}
  J(\boldq)c_{\boldq}s_{\boldq}s_{\boldq^{\prime}}^{2}
  \beta_{\boldq^{\prime}}\beta_{\boldq}^{\dagger}\beta_{\boldq^{\prime}}^{\dagger}\beta_{\boldq},\label{eq:V2}
\end{align}
\begin{align}
  V_{3}=
  &-\frac{2}{N}\sum\limits_{\boldq,\boldq^{\prime}}J(\boldzero)
  (c_{\boldq}^{2}c_{\boldq^{\prime}}^{2}
  \alpha_{\boldq}^{\dagger}\alpha_{\boldq}\beta_{\boldq^{\prime}}^{\dagger}\beta_{\boldq^{\prime}}
  +c_{\boldq}s_{\boldq}c_{\boldq^{\prime}}s_{\boldq^{\prime}}
  \alpha_{\boldq}^{\dagger}\beta_{\boldq}^{\dagger}\alpha_{\boldq^{\prime}}\beta_{\boldq^{\prime}}
  +c_{\boldq}s_{\boldq}c_{\boldq^{\prime}}s_{\boldq^{\prime}}
  \alpha_{\boldq}^{\dagger}\beta_{\boldq}^{\dagger}\beta_{\boldq^{\prime}}^{\dagger}\alpha_{\boldq^{\prime}}^{\dagger}
  \notag\\
  &\ \ \ \ \ \ \ \ \ \ \ \ \ \ \ \ \ \
  +c_{\boldq}s_{\boldq}c_{\boldq^{\prime}}s_{\boldq^{\prime}}
  \beta_{\boldq}\alpha_{\boldq}\alpha_{\boldq^{\prime}}\beta_{\boldq^{\prime}}
  +c_{\boldq}s_{\boldq}c_{\boldq^{\prime}}s_{\boldq^{\prime}}
  \beta_{\boldq}\alpha_{\boldq}\beta_{\boldq^{\prime}}^{\dagger}\alpha_{\boldq^{\prime}}^{\dagger}
  +s_{\boldq}^{2}s_{\boldq^{\prime}}^{2}
  \beta_{\boldq}\beta_{\boldq}^{\dagger}\alpha_{\boldq^{\prime}}\alpha_{\boldq^{\prime}}^{\dagger}
  )\notag\\
  &-\frac{2}{N}\sum\limits_{\boldq,\boldq^{\prime}}J(\boldq-\boldq^{\prime})(
  c_{\boldq}^{2}c_{\boldq^{\prime}}^{2}
  \alpha_{\boldq}^{\dagger}\alpha_{\boldq^{\prime}}\beta_{\boldq}^{\dagger}\beta_{\boldq^{\prime}}
  +c_{\boldq}s_{\boldq}c_{\boldq^{\prime}}s_{\boldq^{\prime}}
  \alpha_{\boldq}^{\dagger}\beta_{\boldq^{\prime}}^{\dagger}\alpha_{\boldq}\beta_{\boldq^{\prime}}
  +c_{\boldq}s_{\boldq}c_{\boldq^{\prime}}s_{\boldq^{\prime}}
  \beta_{\boldq}\alpha_{\boldq^{\prime}}\beta_{\boldq}^{\dagger}\alpha_{\boldq^{\prime}}^{\dagger}
  \notag\\
  &\ \ \ \ \ \ \ \ \ \ \ \ \ \ \ \ \ \ \ \ \ \ \ \ \
  +c_{\boldq}^{2}s_{\boldq^{\prime}}^{2}
  \alpha_{\boldq}^{\dagger}\beta_{\boldq^{\prime}}^{\dagger}\beta_{\boldq}^{\dagger}\alpha_{\boldq^{\prime}}^{\dagger}
  +c_{\boldq}^{2}s_{\boldq^{\prime}}^{2}
  \beta_{\boldq^{\prime}}\alpha_{\boldq}\alpha_{\boldq^{\prime}}\beta_{\boldq}
  +s_{\boldq}^{2}s_{\boldq^{\prime}}^{2}
  \beta_{\boldq}\beta_{\boldq^{\prime}}^{\dagger}\alpha_{\boldq}\alpha_{\boldq^{\prime}}^{\dagger})\notag\\
  &+\frac{2}{N}\sum\limits_{\boldq,\boldq^{\prime}}J(\boldq)(
  c_{\boldq}^{2}c_{\boldq^{\prime}}s_{\boldq^{\prime}}
  \alpha_{\boldq}\alpha_{\boldq^{\prime}}\beta_{\boldq}\beta_{\boldq^{\prime}}
  +c_{\boldq}^{2}c_{\boldq^{\prime}}s_{\boldq^{\prime}}
  \alpha_{\boldq}\beta_{\boldq^{\prime}}^{\dagger}\beta_{\boldq}\alpha_{\boldq^{\prime}}^{\dagger}
  +c_{\boldq^{\prime}}^{2}c_{\boldq}s_{\boldq}
  \alpha_{\boldq}\beta_{\boldq^{\prime}}^{\dagger}\alpha_{\boldq}^{\dagger}\beta_{\boldq^{\prime}}
  \notag\\
  &\ \ \ \ \ \ \ \ \ \ \ \ \ \ \ \ \ \
  +s_{\boldq}^{2}c_{\boldq^{\prime}}s_{\boldq^{\prime}}
  \beta_{\boldq}^{\dagger}\alpha_{\boldq^{\prime}}\alpha_{\boldq}^{\dagger}\beta_{\boldq^{\prime}}
  +s_{\boldq}^{2}c_{\boldq^{\prime}}s_{\boldq^{\prime}}
  \beta_{\boldq}^{\dagger}\beta_{\boldq^{\prime}}^{\dagger}\alpha_{\boldq}^{\dagger}\alpha_{\boldq^{\prime}}^{\dagger}
  +s_{\boldq^{\prime}}^{2}c_{\boldq}s_{\boldq}
  \beta_{\boldq}^{\dagger}\alpha_{\boldq^{\prime}}\beta_{\boldq}\alpha_{\boldq^{\prime}}^{\dagger})\notag\\
  &+\frac{2}{N}\sum\limits_{\boldq,\boldq^{\prime}}J(\boldq)(
  c_{\boldq}^{2}c_{\boldq^{\prime}}s_{\boldq^{\prime}}
  \alpha_{\boldq^{\prime}}^{\dagger}\alpha_{\boldq}\beta_{\boldq^{\prime}}^{\dagger}\beta_{\boldq}
  +c_{\boldq}^{2}c_{\boldq^{\prime}}s_{\boldq^{\prime}}
  \beta_{\boldq^{\prime}}\alpha_{\boldq}\alpha_{\boldq^{\prime}}\beta_{\boldq}
  +c_{\boldq^{\prime}}^{2}c_{\boldq}s_{\boldq}
  \alpha_{\boldq^{\prime}}^{\dagger}\beta_{\boldq}^{\dagger}\alpha_{\boldq^{\prime}}\beta_{\boldq}
  \notag\\
  &\ \ \ \ \ \ \ \ \ \ \ \ \ \ \ \ \ \
  +s_{\boldq}^{2}c_{\boldq^{\prime}}s_{\boldq^{\prime}}
  \alpha_{\boldq^{\prime}}^{\dagger}\beta_{\boldq}^{\dagger}\beta_{\boldq^{\prime}}^{\dagger}\alpha_{\boldq}^{\dagger}
  +s_{\boldq}^{2}c_{\boldq^{\prime}}s_{\boldq^{\prime}}
  \beta_{\boldq^{\prime}}\beta_{\boldq}^{\dagger}\alpha_{\boldq^{\prime}}\alpha_{\boldq}^{\dagger}
  +s_{\boldq^{\prime}}^{2}c_{\boldq}s_{\boldq}  
  \beta_{\boldq^{\prime}}\alpha_{\boldq}\beta_{\boldq^{\prime}}^{\dagger}\alpha_{\boldq}^{\dagger}),\label{eq:V3}\\
  V_{4}=
  &\frac{2}{N}\sum\limits_{\boldq,\boldq^{\prime}}J(\boldzero)(
  s_{\boldq^{\prime}}^{2}c_{\boldq}s_{\boldq}
  \beta_{\boldq}\alpha_{\boldq}\alpha_{\boldq^{\prime}}\alpha_{\boldq^{\prime}}^{\dagger}
  +s_{\boldq^{\prime}}^{2}c_{\boldq}s_{\boldq}
  \alpha_{\boldq}^{\dagger}\beta_{\boldq}^{\dagger}\alpha_{\boldq^{\prime}}\alpha_{\boldq^{\prime}}^{\dagger}
  +c_{\boldq}^{2}c_{\boldq^{\prime}}s_{\boldq^{\prime}}
  \alpha_{\boldq}^{\dagger}\alpha_{\boldq}\beta_{\boldq^{\prime}}^{\dagger}\alpha_{\boldq^{\prime}}^{\dagger}
  +c_{\boldq}^{2}c_{\boldq^{\prime}}s_{\boldq^{\prime}}
  \alpha_{\boldq}^{\dagger}\alpha_{\boldq}\alpha_{\boldq^{\prime}}\beta_{\boldq^{\prime}}
  )\notag\\
  &+\frac{2}{N}\sum\limits_{\boldq,\boldq^{\prime}}J(\boldq-\boldq^{\prime})(
  s_{\boldq^{\prime}}^{2}c_{\boldq}s_{\boldq}
  \beta_{\boldq^{\prime}}\alpha_{\boldq}\alpha_{\boldq^{\prime}}\alpha_{\boldq}^{\dagger}
  +s_{\boldq^{\prime}}^{2}c_{\boldq}s_{\boldq}
  \alpha_{\boldq}^{\dagger}\beta_{\boldq^{\prime}}^{\dagger}\alpha_{\boldq}\alpha_{\boldq^{\prime}}^{\dagger}
  +c_{\boldq}^{2}c_{\boldq^{\prime}}s_{\boldq^{\prime}}
  \alpha_{\boldq}^{\dagger}\alpha_{\boldq^{\prime}}\beta_{\boldq}^{\dagger}\alpha_{\boldq^{\prime}}^{\dagger}
  +c_{\boldq}^{2}c_{\boldq^{\prime}}s_{\boldq^{\prime}}
  \alpha_{\boldq^{\prime}}^{\dagger}\alpha_{\boldq}\alpha_{\boldq^{\prime}}\beta_{\boldq}
  )\notag\\
  &-\frac{2}{N}\sum\limits_{\boldq,\boldq^{\prime}}J(\boldq)(
  s_{\boldq}^{2}s_{\boldq^{\prime}}^{2}
  \beta_{\boldq}^{\dagger}\alpha_{\boldq^{\prime}}\alpha_{\boldq}^{\dagger}\alpha_{\boldq^{\prime}}^{\dagger}
  +c_{\boldq}s_{\boldq}c_{\boldq^{\prime}}s_{\boldq^{\prime}}
  \alpha_{\boldq}\beta_{\boldq^{\prime}}^{\dagger}\alpha_{\boldq}^{\dagger}\alpha_{\boldq^{\prime}}^{\dagger}
  +c_{\boldq}s_{\boldq}c_{\boldq^{\prime}}s_{\boldq^{\prime}}
  \alpha_{\boldq}\alpha_{\boldq^{\prime}}\alpha_{\boldq}^{\dagger}\beta_{\boldq^{\prime}}
  +c_{\boldq}^{2}s_{\boldq^{\prime}}^{2}
  \alpha_{\boldq}\alpha_{\boldq^{\prime}}\beta_{\boldq}\alpha_{\boldq^{\prime}}^{\dagger}
  )\notag\\
  &-\frac{2}{N}\sum\limits_{\boldq,\boldq^{\prime}}J(\boldq)(
    c_{\boldq}s_{\boldq}c_{\boldq^{\prime}}s_{\boldq^{\prime}}
    \beta_{\boldq^{\prime}}\alpha_{\boldq}\alpha_{\boldq^{\prime}}\alpha_{\boldq}^{\dagger}
    +c_{\boldq}s_{\boldq}c_{\boldq^{\prime}}s_{\boldq^{\prime}}
    \alpha_{\boldq^{\prime}}^{\dagger}\alpha_{\boldq}\beta_{\boldq^{\prime}}^{\dagger}\alpha_{\boldq}^{\dagger}
    +s_{\boldq}^{2}c_{\boldq^{\prime}}^{2}
    \alpha_{\boldq^{\prime}}^{\dagger}\beta_{\boldq}^{\dagger}\alpha_{\boldq^{\prime}}\alpha_{\boldq}^{\dagger}
    +c_{\boldq}^{2}c_{\boldq^{\prime}}^{2}
    \alpha_{\boldq^{\prime}}^{\dagger}\alpha_{\boldq}\alpha_{\boldq^{\prime}}\beta_{\boldq}
  ),\label{eq:V4}
\end{align}
and
\begin{align}
  V_{5}=
  &\frac{2}{N}\sum\limits_{\boldq,\boldq^{\prime}}J(\boldzero)(
    c_{\boldq^{\prime}}^{2}c_{\boldq}s_{\boldq}
    \alpha_{\boldq}^{\dagger}\beta_{\boldq}^{\dagger}\beta_{\boldq^{\prime}}^{\dagger}\beta_{\boldq^{\prime}}
    +c_{\boldq^{\prime}}^{2}c_{\boldq}s_{\boldq}
    \beta_{\boldq}\alpha_{\boldq}\beta_{\boldq^{\prime}}^{\dagger}\beta_{\boldq^{\prime}}
    +s_{\boldq}^{2}c_{\boldq^{\prime}}s_{\boldq^{\prime}}
    \beta_{\boldq}\beta_{\boldq}^{\dagger}\alpha_{\boldq^{\prime}}\beta_{\boldq^{\prime}}
    +s_{\boldq}^{2}c_{\boldq^{\prime}}s_{\boldq^{\prime}}
    \beta_{\boldq}\beta_{\boldq}^{\dagger}\beta_{\boldq^{\prime}}^{\dagger}\alpha_{\boldq^{\prime}}^{\dagger}
  )\notag\\
  &+\frac{2}{N}\sum\limits_{\boldq,\boldq^{\prime}}J(\boldq-\boldq^{\prime})(
    c_{\boldq}^{2}c_{\boldq^{\prime}}s_{\boldq^{\prime}}
    \alpha_{\boldq}^{\dagger}\beta_{\boldq^{\prime}}^{\dagger}\beta_{\boldq}^{\dagger}\beta_{\boldq^{\prime}}
    +c_{\boldq^{\prime}}^{2}c_{\boldq}s_{\boldq}
    \beta_{\boldq}\alpha_{\boldq^{\prime}}\beta_{\boldq}^{\dagger}\beta_{\boldq^{\prime}}
    +s_{\boldq}^{2}c_{\boldq^{\prime}}s_{\boldq^{\prime}}
    \beta_{\boldq}\beta_{\boldq^{\prime}}^{\dagger}\alpha_{\boldq}\beta_{\boldq^{\prime}}
    +s_{\boldq^{\prime}}^{2}c_{\boldq}s_{\boldq}
    \beta_{\boldq}\beta_{\boldq^{\prime}}^{\dagger}\beta_{\boldq}^{\dagger}\alpha_{\boldq^{\prime}}^{\dagger}
  )\notag\\
  &-\frac{2}{N}\sum\limits_{\boldq,\boldq^{\prime}}J(\boldq)(
    c_{\boldq}^{2}c_{\boldq^{\prime}}^{2}
    \alpha_{\boldq}\beta_{\boldq^{\prime}}^{\dagger}\beta_{\boldq}\beta_{\boldq^{\prime}}
    +c_{\boldq}s_{\boldq}c_{\boldq^{\prime}}s_{\boldq^{\prime}}
    \beta_{\boldq}^{\dagger}\alpha_{\boldq^{\prime}}\beta_{\boldq}\beta_{\boldq^{\prime}}
    +c_{\boldq}s_{\boldq}c_{\boldq^{\prime}}s_{\boldq^{\prime}}
    \beta_{\boldq}^{\dagger}\beta_{\boldq^{\prime}}^{\dagger}\beta_{\boldq}\alpha_{\boldq^{\prime}}^{\dagger}
    +s_{\boldq}^{2}c_{\boldq^{\prime}}^{2}
    \beta_{\boldq}^{\dagger}\beta_{\boldq^{\prime}}^{\dagger}\alpha_{\boldq}^{\dagger}\beta_{\boldq^{\prime}}
  )\notag\\
  &-\frac{2}{N}\sum\limits_{\boldq,\boldq^{\prime}}J(\boldq)(
    c_{\boldq}s_{\boldq}c_{\boldq^{\prime}}s_{\boldq^{\prime}}
    \alpha_{\boldq^{\prime}}^{\dagger}\beta_{\boldq}^{\dagger}\beta_{\boldq^{\prime}}^{\dagger}\beta_{\boldq}
    +c_{\boldq}^{2}s_{\boldq^{\prime}}^{2}
    \beta_{\boldq^{\prime}}\alpha_{\boldq}\beta_{\boldq^{\prime}}^{\dagger}\beta_{\boldq}
    +c_{\boldq}s_{\boldq}c_{\boldq^{\prime}}s_{\boldq^{\prime}}
    \beta_{\boldq^{\prime}}\beta_{\boldq}^{\dagger}\alpha_{\boldq^{\prime}}\beta_{\boldq}
    +s_{\boldq}^{2}s_{\boldq^{\prime}}^{2}
    \beta_{\boldq^{\prime}}\beta_{\boldq}^{\dagger}\beta_{\boldq^{\prime}}^{\dagger}\alpha_{\boldq}^{\dagger}
  ).\label{eq:V5}
\end{align}
Therefore
$H_{\textrm{intra}}$ is given by 
$H_{\textrm{intra}}=V_{1}+V_{2}+(\textrm{H.c.})$
with Eqs. (\ref{eq:V1}) and (\ref{eq:V2}),
and
$H_{\textrm{inter}}$ is given by
$H_{\textrm{inter}}=V_{3}+V_{4}+V_{5}+(\textrm{H.c.})$
with Eqs. (\ref{eq:V3}){--}(\ref{eq:V5}). 
Note that each term in Eqs. (\ref{eq:V1}){--}(\ref{eq:V5})
consists of the terms for $\boldq=\boldq^{\prime}=\boldzero$,
for $\boldq=\boldzero$, $\boldq^{\prime}\neq\boldzero$,
for $\boldq\neq\boldzero$, $\boldq^{\prime}=\boldzero$,
and for $\boldq\neq\boldzero$, $\boldq^{\prime}\neq\boldzero$.

\section{Derivation of
  the mean-field interaction terms of
  $H_{\textrm{intra}}$ and $H_{\textrm{inter}}$
  for $\boldq=\boldq^{\prime}=\boldzero$,
  for $\boldq=\boldzero$, $\boldq^{\prime}\neq\boldzero$,
  and for $\boldq\neq\boldzero$, $\boldq^{\prime}=\boldzero$
  and leading-order estimate}

We derive the mean-field interaction terms which arise from
the terms of $H_{\textrm{int}}=H_{\textrm{intra}}+H_{\textrm{inter}}$
for $\boldq=\boldq^{\prime}=\boldzero$,
for $\boldq=\boldzero$, $\boldq^{\prime}\neq\boldzero$,
and for $\boldq\neq\boldzero$, $\boldq^{\prime}=\boldzero$,
and we estimate the coefficients within the leading order.
This derivation can be performed in a manner similar to that of 
the Bogoliubov theory~\cite{Bogoliubov,AGD} for the BEC in a Bose gas.
Since the condensed states in the BEC of magnons
in the antiferromagnet without the external magnetic field
are the $\boldq=\boldzero$ states of the $\alpha$- and $\beta$-band magnons, 
we can use the following Bogoliubov approximations:
$\alpha_{\boldzero}\approx \sqrt{n_{\boldzero\alpha}}$,
$\alpha_{\boldzero}^{\dagger}\approx \sqrt{n_{\boldzero\alpha}}$,
$\beta_{\boldzero}\approx \sqrt{n_{\boldzero\beta}}$, and 
$\beta_{\boldzero}^{\dagger}\approx \sqrt{n_{\boldzero\beta}}$.
Note that 
$\sqrt{n_{\boldzero\alpha}}$ and $\sqrt{n_{\boldzero\beta}}$ satisfy 
$\sqrt{n_{\boldzero\alpha}}=\sqrt{N_{\alpha}-\sum_{\boldq\neq\boldzero}\alpha_{\boldq}^{\dagger}\alpha_{\boldq}}$
and
$\sqrt{n_{\boldzero\beta}}=\sqrt{N_{\beta}-\sum_{\boldq\neq\boldzero}\beta_{\boldq}^{\dagger}\beta_{\boldq}}$, 
where $N_{\alpha}$ and $N_{\beta}$ are the numbers of the $\alpha$- and $\beta$-band magnons 
($N_{\alpha}=N_{\beta}\equiv N_{\textrm{m}}$ is satisfied 
in the absence of the external magnetic field).
Because of the Bogoliubov approximations, for example
the first term of $V_{1}$ in Eq. (\ref{eq:V1}) gives the following mean-field interaction terms:
\begin{align}
  -\frac{2}{N}
  J(\boldzero)c_{\boldzero}^{2}s_{\boldzero}^{2}n_{\boldzero\alpha}^{2}
  -\frac{2}{N}\sum\limits_{\boldq^{\prime}\neq \boldzero}
  J(\boldzero)c_{\boldzero}^{2}s_{\boldq^{\prime}}^{2}
  n_{\boldzero\alpha}\alpha_{\boldq^{\prime}}\alpha_{\boldq^{\prime}}^{\dagger}
  -\frac{2}{N}\sum\limits_{\boldq\neq \boldzero}
  J(\boldzero)c_{\boldq}^{2}s_{\boldzero}^{2}
  n_{\boldzero\alpha}\alpha_{\boldq}^{\dagger}\alpha_{\boldq}.\label{eq:ex}
\end{align}
(Here we have neglected the contribution for $\boldq\neq\boldzero$, $\boldq^{\prime}\neq\boldzero$.) 
In a similar manner
we can derive the mean-field interaction terms 
for $\boldq=\boldq^{\prime}=\boldzero$,
for $\boldq=\boldzero$, $\boldq^{\prime}\neq\boldzero$,
and for $\boldq\neq\boldzero$, $\boldq^{\prime}=\boldzero$
from the other terms of $V_{1}$ and the terms of $V_{2}$, $V_{3}$, $V_{4}$, and $V_{5}$.
By performing a similar calculation for each term  
in Eqs. (\ref{eq:V1}){--}(\ref{eq:V5}) 
and combining these results with
$H_{\alpha\alpha\alpha\alpha}=V_{1}+V_{1}^{\dagger}$, $H_{\beta\beta\beta\beta}=V_{2}+V_{2}^{\dagger}$,
$H_{\alpha\alpha\beta\beta}=V_{3}+V_{3}^{\dagger}$, $H_{\alpha\alpha\alpha\beta}=V_{4}+V_{4}^{\dagger}$,
and $H_{\alpha\beta\beta\beta}=V_{5}+V_{5}^{\dagger}$,
we can write the mean-field interaction terms
which arise from the terms for $\boldq=\boldq^{\prime}=\boldzero$,
for $\boldq=\boldzero$, $\boldq^{\prime}\neq\boldzero$,
and for $\boldq\neq\boldzero$, $\boldq^{\prime}=\boldzero$
as $H_{\alpha\alpha\alpha\alpha}^{(0)}$, $H_{\beta\beta\beta\beta}^{(0)}$,
$H_{\alpha\alpha\beta\beta}^{(0)}$, $H_{\alpha\alpha\alpha\beta}^{(0)}$,
and $H_{\alpha\beta\beta\beta}^{(0)}$,
where 
\begin{align}
  H_{\alpha\alpha\alpha\alpha}^{(0)}
  =&\frac{4}{N}N_{\textrm{m}}^{2}J(\boldzero)c_{\boldzero}s_{\boldzero}
  (c_{\boldzero}^{2}+s_{\boldzero}^{2}-2c_{\boldzero}s_{\boldzero})
  -\frac{4}{N}N_{\textrm{m}}J(\boldzero)c_{\boldzero}s_{\boldzero}
  (c_{\boldzero}^{2}+s_{\boldzero}^{2}-2c_{\boldzero}s_{\boldzero})
  \sum\limits_{\boldq\neq \boldzero}\alpha_{\boldq}^{\dagger}\alpha_{\boldq}\notag\\
  &-\frac{4}{N}N_{\textrm{m}}
  \sum\limits_{\boldq\neq \boldzero}\alpha_{\boldq}^{\dagger}\alpha_{\boldq}
  [J(\boldzero)(c_{\boldq}^{2}s_{\boldzero}^{2}+c_{\boldzero}^{2}s_{\boldq}^{2})
    +2J(\boldq)c_{\boldq}s_{\boldq}c_{\boldzero}s_{\boldzero}
    -J(\boldzero)c_{\boldzero}s_{\boldzero}(c_{\boldq}^{2}+s_{\boldq}^{2})
    -J(\boldq)c_{\boldq}s_{\boldq}(c_{\boldzero}^{2}+s_{\boldzero}^{2})],\label{eq:Hintra1}\\
  H_{\beta\beta\beta\beta}^{(0)}
  =&\frac{4}{N}N_{\textrm{m}}^{2}J(\boldzero)c_{\boldzero}s_{\boldzero}
  (c_{\boldzero}^{2}+s_{\boldzero}^{2}-2c_{\boldzero}s_{\boldzero})
  -\frac{4}{N}N_{\textrm{m}}J(\boldzero)c_{\boldzero}s_{\boldzero}
  (c_{\boldzero}^{2}+s_{\boldzero}^{2}-2c_{\boldzero}s_{\boldzero})
  \sum\limits_{\boldq\neq \boldzero}\beta_{\boldq}^{\dagger}\beta_{\boldq}\notag\\
  &-\frac{4}{N}N_{\textrm{m}}
  \sum\limits_{\boldq\neq \boldzero}\beta_{\boldq}^{\dagger}\beta_{\boldq}
  [J(\boldzero)(c_{\boldq}^{2}s_{\boldzero}^{2}+c_{\boldzero}^{2}s_{\boldq}^{2})
    +2J(\boldq)c_{\boldq}s_{\boldq}c_{\boldzero}s_{\boldzero}
    -J(\boldzero)c_{\boldzero}s_{\boldzero}(c_{\boldq}^{2}+s_{\boldq}^{2})
    -J(\boldq)c_{\boldq}s_{\boldq}(c_{\boldzero}^{2}+s_{\boldzero}^{2})],\label{eq:Hintra2}\\
  H_{\alpha\alpha\beta\beta}^{(0)}
  =&-\frac{8}{N}N_{\textrm{m}}^{2}J(\boldzero)
  (c_{\boldzero}^{2}+s_{\boldzero}^{2}-c_{\boldzero}s_{\boldzero})
  (c_{\boldzero}^{2}+s_{\boldzero}^{2}-2c_{\boldzero}s_{\boldzero})\notag\\
  &+\frac{8}{N}N_{\textrm{m}}J(\boldzero)
  (c_{\boldzero}^{2}+s_{\boldzero}^{2}-c_{\boldzero}s_{\boldzero})
  (c_{\boldzero}^{2}+s_{\boldzero}^{2}-2c_{\boldzero}s_{\boldzero})
  \sum\limits_{\boldq\neq \boldzero}
  (\alpha_{\boldq}^{\dagger}\alpha_{\boldq}+\beta_{\boldq}^{\dagger}\beta_{\boldq})\notag\\
  &-\frac{4}{N}N_{\textrm{m}}
  \sum\limits_{\boldq\neq \boldzero}
  (\alpha_{\boldq}^{\dagger}\alpha_{\boldq}+\beta_{\boldq}^{\dagger}\beta_{\boldq})
  \{
  J(\boldzero)[(c_{\boldq}^{2}c_{\boldzero}^{2}+s_{\boldq}^{2}s_{\boldzero}^{2})
    -c_{\boldzero}s_{\boldzero}(c_{\boldq}^{2}+s_{\boldq}^{2})]
  -J(\boldq)c_{\boldq}s_{\boldq}
  (c_{\boldzero}^{2}+s_{\boldzero}^{2}-2c_{\boldzero}s_{\boldzero})\}\notag\\
  &+\frac{4}{N}N_{\textrm{m}}
  \sum\limits_{\boldq\neq \boldzero}
  (\alpha_{\boldq}\beta_{\boldq}+\alpha_{\boldq}^{\dagger}\beta_{\boldq}^{\dagger})
  (c_{\boldzero}^{2}+s_{\boldzero}^{2}-2c_{\boldzero}s_{\boldzero})
  [2J(\boldzero)c_{\boldq}s_{\boldq}-J(\boldq)(c_{\boldq}^{2}+s_{\boldq}^{2})],\label{eq:Hinter1}          
\end{align}
\begin{align}
  H_{\alpha\alpha\alpha\beta}^{(0)}
  =&-\frac{4}{N}N_{\textrm{m}}^{2}J(\boldzero)
  +\frac{2}{N}N_{\textrm{m}}J(\boldzero)\sum\limits_{\boldq\neq \boldzero}
  (3\alpha_{\boldq}^{\dagger}\alpha_{\boldq}+\beta_{\boldq}^{\dagger}\beta_{\boldq})\notag\\
  &+\frac{16}{N}N_{\textrm{m}}^{2}J(\boldzero)c_{\boldzero}s_{\boldzero}
  (c_{\boldzero}^{2}+s_{\boldzero}^{2}-2c_{\boldzero}s_{\boldzero})
  -\frac{8}{N}N_{\textrm{m}}J(\boldzero)c_{\boldzero}s_{\boldzero}
  (c_{\boldzero}^{2}+s_{\boldzero}^{2}-2c_{\boldzero}s_{\boldzero})
  \sum\limits_{\boldq\neq \boldzero}
  (3\alpha_{\boldq}^{\dagger}\alpha_{\boldq}+\beta_{\boldq}^{\dagger}\beta_{\boldq})\notag\\
  &-\frac{4}{N}N_{\textrm{m}}
  (c_{\boldzero}^{2}+s_{\boldzero}^{2}-2c_{\boldzero}s_{\boldzero})
  \sum\limits_{\boldq\neq \boldzero}\alpha_{\boldq}^{\dagger}\alpha_{\boldq}
  [J(\boldzero)(c_{\boldq}^{2}+s_{\boldq}^{2})-2J(\boldq)c_{\boldq}s_{\boldq}]\notag\\
  &+\frac{2}{N}N_{\textrm{m}}
  \sum\limits_{\boldq\neq \boldzero}
  (\alpha_{\boldq}\beta_{\boldq}+\alpha_{\boldq}^{\dagger}\beta_{\boldq}^{\dagger})
  (c_{\boldzero}^{2}+s_{\boldzero}^{2}-2c_{\boldzero}s_{\boldzero})
  [2J(\boldzero)c_{\boldq}s_{\boldq}-J(\boldq)(c_{\boldq}^{2}+s_{\boldq}^{2})],\label{eq:Hinter2}
\end{align}
and 
\begin{align}
  H_{\alpha\beta\beta\beta}^{(0)}
  =&-\frac{4}{N}N_{\textrm{m}}^{2}J(\boldzero)
  +\frac{2}{N}N_{\textrm{m}}J(\boldzero)\sum\limits_{\boldq\neq \boldzero}
  (\alpha_{\boldq}^{\dagger}\alpha_{\boldq}+3\beta_{\boldq}^{\dagger}\beta_{\boldq})\notag\\
  &+\frac{16}{N}N_{\textrm{m}}^{2}J(\boldzero)c_{\boldzero}s_{\boldzero}
  (c_{\boldzero}^{2}+s_{\boldzero}^{2}-2c_{\boldzero}s_{\boldzero})
  -\frac{8}{N}N_{\textrm{m}}J(\boldzero)c_{\boldzero}s_{\boldzero}
  (c_{\boldzero}^{2}+s_{\boldzero}^{2}-2c_{\boldzero}s_{\boldzero})
  \sum\limits_{\boldq\neq \boldzero}
  (\alpha_{\boldq}^{\dagger}\alpha_{\boldq}+3\beta_{\boldq}^{\dagger}\beta_{\boldq})\notag\\
  &-\frac{4}{N}N_{\textrm{m}}
  (c_{\boldzero}^{2}+s_{\boldzero}^{2}-2c_{\boldzero}s_{\boldzero})
  \sum\limits_{\boldq\neq \boldzero}\beta_{\boldq}^{\dagger}\beta_{\boldq}
  [J(\boldzero)(c_{\boldq}^{2}+s_{\boldq}^{2})-2J(\boldq)c_{\boldq}s_{\boldq}]\notag\\
  &+\frac{2}{N}N_{\textrm{m}}
  \sum\limits_{\boldq\neq \boldzero}
  (\alpha_{\boldq}\beta_{\boldq}+\alpha_{\boldq}^{\dagger}\beta_{\boldq}^{\dagger})
  (c_{\boldzero}^{2}+s_{\boldzero}^{2}-2c_{\boldzero}s_{\boldzero})
  [2J(\boldzero)c_{\boldq}s_{\boldq}-J(\boldq)(c_{\boldq}^{2}+s_{\boldq}^{2})].\label{eq:Hinter3}
\end{align} 
We have used 
\begin{align}
  &n_{\boldzero\alpha}^{2}
  =(N_{\alpha}-\sum\limits_{\boldq\neq\boldzero}\alpha_{\boldq}^{\dagger}\alpha_{\boldq})^{2}
  \approx
  N_{\alpha}^{2}-2N_{\alpha}\sum\limits_{\boldq\neq\boldzero}\alpha_{\boldq}^{\dagger}\alpha_{\boldq}
  =
  N_{\textrm{m}}^{2}-2N_{\textrm{m}}\sum\limits_{\boldq\neq\boldzero}\alpha_{\boldq}^{\dagger}\alpha_{\boldq},\\
  &n_{\boldzero\alpha}\sum\limits_{\boldq\neq\boldzero}\alpha_{\boldq}^{\dagger}\alpha_{\boldq}
  \approx 
  N_{\alpha}\sum\limits_{\boldq\neq\boldzero}\alpha_{\boldq}^{\dagger}\alpha_{\boldq}
  =
  N_{\textrm{m}}\sum\limits_{\boldq\neq\boldzero}\alpha_{\boldq}^{\dagger}\alpha_{\boldq},
\end{align}
for the derivation of Eq. (\ref{eq:Hintra1}),
\begin{align}
  &n_{\boldzero\beta}^{2}
  =(N_{\beta}-\sum\limits_{\boldq\neq\boldzero}\beta_{\boldq}^{\dagger}\beta_{\boldq})^{2}
  \approx
  N_{\beta}^{2}-2N_{\beta}\sum\limits_{\boldq\neq\boldzero}\beta_{\boldq}^{\dagger}\beta_{\boldq}
  =
  N_{\textrm{m}}^{2}-2N_{\textrm{m}}\sum\limits_{\boldq\neq\boldzero}\beta_{\boldq}^{\dagger}\beta_{\boldq},\\
  &n_{\boldzero\beta}\sum\limits_{\boldq\neq\boldzero}\beta_{\boldq}^{\dagger}\beta_{\boldq}
  \approx 
  N_{\beta}\sum\limits_{\boldq\neq\boldzero}\beta_{\boldq}^{\dagger}\beta_{\boldq}
  =N_{\textrm{m}}\sum\limits_{\boldq\neq\boldzero}\beta_{\boldq}^{\dagger}\beta_{\boldq},
\end{align}
for the derivation of Eq. (\ref{eq:Hintra2}),
\begin{align}
  &n_{\boldzero\alpha}n_{\boldzero\beta}
  \approx 
  N_{\alpha}N_{\beta}
  -N_{\alpha}\sum\limits_{\boldq\neq\boldzero}\beta_{\boldq}^{\dagger}\beta_{\boldq}
  -N_{\beta}\sum\limits_{\boldq\neq\boldzero}\alpha_{\boldq}^{\dagger}\alpha_{\boldq}
  = N_{\textrm{m}}^{2}
  -N_{\textrm{m}}\sum\limits_{\boldq\neq\boldzero}
  (\alpha_{\boldq}^{\dagger}\alpha_{\boldq}+\beta_{\boldq}^{\dagger}\beta_{\boldq}),\\
  &n_{\boldzero\beta}\sum\limits_{\boldq\neq\boldzero}\alpha_{\boldq}^{\dagger}\alpha_{\boldq}
  \approx 
  N_{\beta}\sum\limits_{\boldq\neq\boldzero}\alpha_{\boldq}^{\dagger}\alpha_{\boldq}
  =
  N_{\textrm{m}}\sum\limits_{\boldq\neq\boldzero}\alpha_{\boldq}^{\dagger}\alpha_{\boldq},\\
  &n_{\boldzero\alpha}\sum\limits_{\boldq\neq\boldzero}\beta_{\boldq}^{\dagger}\beta_{\boldq}
  \approx 
  N_{\alpha}\sum\limits_{\boldq\neq\boldzero}\beta_{\boldq}^{\dagger}\beta_{\boldq}
  =
  N_{\textrm{m}}\sum\limits_{\boldq\neq\boldzero}\beta_{\boldq}^{\dagger}\beta_{\boldq},\\
  &\sqrt{n_{\boldzero\alpha}}\sqrt{n_{\boldzero\beta}}
  \sum\limits_{\boldq\neq \boldzero}
  (\alpha_{\boldq}\beta_{\boldq}+\alpha_{\boldq}^{\dagger}\beta_{\boldq}^{\dagger})
  \approx 
  \sqrt{N_{\alpha}N_{\beta}}
  \sum\limits_{\boldq\neq \boldzero}
  (\alpha_{\boldq}\beta_{\boldq}+\alpha_{\boldq}^{\dagger}\beta_{\boldq}^{\dagger})
  =N_{\textrm{m}}
  \sum\limits_{\boldq\neq \boldzero}
  (\alpha_{\boldq}\beta_{\boldq}+\alpha_{\boldq}^{\dagger}\beta_{\boldq}^{\dagger}),
\end{align}
for the derivation of Eq. (\ref{eq:Hinter1}),
\begin{align}
  n_{\boldzero\alpha}\sqrt{n_{\boldzero\alpha}}\sqrt{n_{\boldzero\beta}}
  \approx &
  (N_{\alpha}-\sum\limits_{\boldq\neq\boldzero}\alpha_{\boldq}^{\dagger}\alpha_{\boldq})
  [\sqrt{N_{\alpha}N_{\beta}}
    -\frac{1}{2}\sum\limits_{\boldq\neq\boldzero}
    (\alpha_{\boldq}^{\dagger}\alpha_{\boldq}+\beta_{\boldq}^{\dagger}\beta_{\boldq})]\notag\\
  \approx &
  N_{\alpha}\sqrt{N_{\alpha}N_{\beta}}
  -\sqrt{N_{\alpha}N_{\beta}}\sum\limits_{\boldq\neq\boldzero}\alpha_{\boldq}^{\dagger}\alpha_{\boldq}
  -\frac{1}{2}N_{\alpha}\sum\limits_{\boldq\neq\boldzero}
  (\alpha_{\boldq}^{\dagger}\alpha_{\boldq}+\beta_{\boldq}^{\dagger}\beta_{\boldq})\notag\\
  =&
  N_{\textrm{m}}^{2}
  -\frac{1}{2}N_{\textrm{m}}\sum\limits_{\boldq\neq\boldzero}
  (3\alpha_{\boldq}^{\dagger}\alpha_{\boldq}+\beta_{\boldq}^{\dagger}\beta_{\boldq}),\\
  \sqrt{n_{\boldzero\alpha}}\sqrt{n_{\boldzero\beta}}
  \sum\limits_{\boldq\neq\boldzero}\alpha_{\boldq}^{\dagger}\alpha_{\boldq}
  \approx &
  \sqrt{N_{\alpha}N_{\beta}}
  \sum\limits_{\boldq\neq\boldzero}\alpha_{\boldq}^{\dagger}\alpha_{\boldq}
  =N_{\textrm{m}}
  \sum\limits_{\boldq\neq\boldzero}\alpha_{\boldq}^{\dagger}\alpha_{\boldq},\\
  n_{\boldzero\alpha}
  \sum\limits_{\boldq\neq \boldzero}
  (\alpha_{\boldq}\beta_{\boldq}+\alpha_{\boldq}^{\dagger}\beta_{\boldq}^{\dagger})
  \approx &
  N_{\alpha}\sum\limits_{\boldq\neq \boldzero}
  (\alpha_{\boldq}\beta_{\boldq}+\alpha_{\boldq}^{\dagger}\beta_{\boldq}^{\dagger})
  =
  N_{\textrm{m}}\sum\limits_{\boldq\neq \boldzero}
  (\alpha_{\boldq}\beta_{\boldq}+\alpha_{\boldq}^{\dagger}\beta_{\boldq}^{\dagger}),
\end{align}
for the derivation of Eq. (\ref{eq:Hinter2}), and
\begin{align}
  n_{\boldzero\beta}\sqrt{n_{\boldzero\alpha}}\sqrt{n_{\boldzero\beta}}
  \approx &
  (N_{\beta}-\sum\limits_{\boldq\neq\boldzero}\beta_{\boldq}^{\dagger}\beta_{\boldq})
  [\sqrt{N_{\alpha}N_{\beta}}
    -\frac{1}{2}\sum\limits_{\boldq\neq\boldzero}
    (\alpha_{\boldq}^{\dagger}\alpha_{\boldq}+\beta_{\boldq}^{\dagger}\beta_{\boldq})]\notag\\
  \approx &
  N_{\beta}\sqrt{N_{\alpha}N_{\beta}}
  -\sqrt{N_{\alpha}N_{\beta}}\sum\limits_{\boldq\neq\boldzero}\beta_{\boldq}^{\dagger}\beta_{\boldq}
  -\frac{1}{2}N_{\beta}\sum\limits_{\boldq\neq\boldzero}
  (\alpha_{\boldq}^{\dagger}\alpha_{\boldq}+\beta_{\boldq}^{\dagger}\beta_{\boldq})\notag\\
  =&
  N_{\textrm{m}}^{2}
  -\frac{1}{2}N_{\textrm{m}}\sum\limits_{\boldq\neq\boldzero}
  (\alpha_{\boldq}^{\dagger}\alpha_{\boldq}+3\beta_{\boldq}^{\dagger}\beta_{\boldq}),\\
  \sqrt{n_{\boldzero\alpha}}\sqrt{n_{\boldzero\beta}}
  \sum\limits_{\boldq\neq\boldzero}\beta_{\boldq}^{\dagger}\beta_{\boldq}
  \approx &
  \sqrt{N_{\alpha}N_{\beta}}
  \sum\limits_{\boldq\neq\boldzero}\beta_{\boldq}^{\dagger}\beta_{\boldq}
  =N_{\textrm{m}}
  \sum\limits_{\boldq\neq\boldzero}\beta_{\boldq}^{\dagger}\beta_{\boldq},\\
  n_{\boldzero\beta}
  \sum\limits_{\boldq\neq \boldzero}
  (\alpha_{\boldq}\beta_{\boldq}+\alpha_{\boldq}^{\dagger}\beta_{\boldq}^{\dagger})
  \approx &
  N_{\beta}\sum\limits_{\boldq\neq \boldzero}
  (\alpha_{\boldq}\beta_{\boldq}+\alpha_{\boldq}^{\dagger}\beta_{\boldq}^{\dagger})
  =
  N_{\textrm{m}}\sum\limits_{\boldq\neq \boldzero}
  (\alpha_{\boldq}\beta_{\boldq}+\alpha_{\boldq}^{\dagger}\beta_{\boldq}^{\dagger}),
\end{align}
for the derivation of Eq. (\ref{eq:Hinter3}).
(As described above, $N_{\alpha}=N_{\beta}\equiv N_{\textrm{m}}$ for $h=0$.) 
Since the operator terms in Eqs. (\ref{eq:Hintra1}){--}(\ref{eq:Hinter3})
are relevant to the interaction effects
on the stability of the magnon BEC,
we consider only the operator parts in the following.
Therefore we have
\begin{align}
  &H_{\alpha\alpha\alpha\alpha}^{(0)}+H_{\beta\beta\beta\beta}^{(0)}
  =\frac{N_{\textrm{m}}}{N}\sum\limits_{\boldq\neq \boldzero}
  \Gamma_{\textrm{intra}}^{(0)}(\boldq)
  (\alpha_{\boldq}^{\dagger}\alpha_{\boldq}+\beta_{\boldq}^{\dagger}\beta_{\boldq}),\\
  &H_{\alpha\alpha\beta\beta}^{(0)}
  =\frac{N_{\textrm{m}}}{N}\sum\limits_{\boldq\neq \boldzero}
  \Gamma_{\textrm{inter}1}^{(0)}(\boldq)
  (\alpha_{\boldq}^{\dagger}\alpha_{\boldq}+\beta_{\boldq}^{\dagger}\beta_{\boldq})
  +\frac{N_{\textrm{m}}}{N}\sum\limits_{\boldq\neq \boldzero}
  \Gamma_{\textrm{inter}2}^{(0)}(\boldq)
  (\alpha_{\boldq}\beta_{\boldq}+\alpha_{\boldq}^{\dagger}\beta_{\boldq}^{\dagger}),\\
  &H_{\alpha\alpha\alpha\beta}^{(0)}+H_{\alpha\beta\beta\beta}^{(0)}
  =\frac{N_{\textrm{m}}}{N}\sum\limits_{\boldq\neq \boldzero}
  \Gamma_{\textrm{inter}3}^{(0)}(\boldq)
  (\alpha_{\boldq}^{\dagger}\alpha_{\boldq}+\beta_{\boldq}^{\dagger}\beta_{\boldq})
  +\frac{N_{\textrm{m}}}{N}\sum\limits_{\boldq\neq \boldzero}
  \Gamma_{\textrm{inter}2}^{(0)}(\boldq)
  (\alpha_{\boldq}\beta_{\boldq}+\alpha_{\boldq}^{\dagger}\beta_{\boldq}^{\dagger}).
\end{align}
Here we have introduced the following quantities: 
\begin{align}
  \Gamma_{\textrm{intra}}^{(0)}(\boldq)
  =&-8J(\boldzero)c_{\boldzero}s_{\boldzero}(c_{\boldzero}^{2}+s_{\boldzero}^{2}-2c_{\boldzero}s_{\boldzero})
  -4J(\boldzero)(c_{\boldq}^{2}s_{\boldzero}^{2}+c_{\boldzero}^{2}s_{\boldq}^{2})
  -8J(\boldq)c_{\boldq}s_{\boldq}c_{\boldzero}s_{\boldzero}
  +4J(\boldzero)c_{\boldzero}s_{\boldzero}(c_{\boldq}^{2}+s_{\boldq}^{2})\notag\\
  &+4J(\boldq)c_{\boldq}s_{\boldq}(c_{\boldzero}^{2}+s_{\boldzero}^{2}),\label{eq:Gam1}\\
  \Gamma_{\textrm{inter}1}^{(0)}(\boldq)
  =&8J(\boldzero)(c_{\boldzero}^{2}+s_{\boldzero}^{2}-c_{\boldzero}s_{\boldzero})
  (c_{\boldzero}^{2}+s_{\boldzero}^{2}-2c_{\boldzero}s_{\boldzero})
  -4J(\boldzero)
  [(c_{\boldq}^{2}c_{\boldzero}^{2}+s_{\boldq}^{2}s_{\boldzero}^{2})
    -c_{\boldzero}s_{\boldzero}(c_{\boldq}^{2}+s_{\boldq}^{2})]\notag\\
  &+4J(\boldq)c_{\boldq}s_{\boldq}
  (c_{\boldzero}^{2}+s_{\boldzero}^{2}-2c_{\boldzero}s_{\boldzero}),\label{eq:Gam2}\\
  \Gamma_{\textrm{inter}2}^{(0)}(\boldq)
  =&-4(c_{\boldzero}^{2}+s_{\boldzero}^{2}-2c_{\boldzero}s_{\boldzero})
  [J(\boldq)(c_{\boldq}^{2}+s_{\boldq}^{2})-2J(\boldzero)c_{\boldq}s_{\boldq}],\label{eq:Gam3}\\
  \Gamma_{\textrm{inter}3}^{(0)}(\boldq)
  =&8J(\boldzero)
  -4(c_{\boldzero}^{2}+s_{\boldzero}^{2}-2c_{\boldzero}s_{\boldzero})
  [8J(\boldzero)c_{\boldzero}s_{\boldzero}+J(\boldzero)(c_{\boldq}^{2}+s_{\boldq}^{2})
    -2J(\boldq)c_{\boldq}s_{\boldq}].\label{eq:Gam4}
\end{align}

Since the low-$\boldq$ magnons of the $\alpha$ and $\beta$ bands
describe the low-energy noncondensates in the magnon BEC
of the antiferromagnet for $h=0$,
we estimate $\Gamma_{\textrm{intra}}^{(0)}(\boldq)$,
$\Gamma_{\textrm{inter}1}^{(0)}(\boldq)$,
$\Gamma_{\textrm{inter}2}^{(0)}(\boldq)$, and 
$\Gamma_{\textrm{inter}3}^{(0)}(\boldq)$ in the limit $|\boldq|\rightarrow 0$.
To estimate them,
we use the following identities:
\begin{align}
  &c_{\boldq}^{2}-s_{\boldq}^{2}=1,\label{eq:iden1}\\
  &c_{\boldq}^{2}+s_{\boldq}^{2}=\cosh2\theta_{\boldq},\label{eq:iden2}\\
  &2c_{\boldq}s_{\boldq}=\sinh2\theta_{\boldq}.\label{eq:iden3}
\end{align} 
In addition,
$\cosh2\theta_{\boldq}$ and $\sinh2\theta_{\boldq}$
are given for our antiferromagnet by
\begin{align}
  \cosh 2\theta_{\boldq} =
  \frac{J(\boldzero)+K}{\sqrt{(J(\boldzero)+K)^{2}-J(\boldq)^{2}}}
  \sim \frac{J(\boldzero)+K}{\sqrt{2J(\boldzero)K}},\label{eq:cosh2}\\
  \sinh 2\theta_{\boldq} =
  \frac{J(\boldq)}{\sqrt{(J(\boldzero)+K)^{2}-J(\boldq)^{2}}}
  \sim \frac{J(\boldzero)}{\sqrt{2J(\boldzero)K}}.\label{eq:sinh2} 
\end{align}
(In the above limiting expressions
we have considered only the leading terms  in the limit $|\boldq|\rightarrow 0$ for $J\gg K$.)
By using Eqs. (\ref{eq:iden1}){--}(\ref{eq:sinh2})
and retaining the leading terms in the limit $|\boldq|\rightarrow 0$ for $J\gg K$,
we obtain
\begin{align}
  &-8J(\boldzero)c_{\boldzero}s_{\boldzero}(c_{\boldzero}^{2}+s_{\boldzero}^{2}-2c_{\boldzero}s_{\boldzero})
  \sim -4J(\boldzero)\sqrt{\frac{K}{2J(\boldzero)}}\sqrt{\frac{J(\boldzero)}{2K}}
  =-2J(\boldzero),\label{eq:First}\\
  &-4J(\boldzero)(c_{\boldq}^{2}s_{\boldzero}^{2}+c_{\boldzero}^{2}s_{\boldq}^{2})
  +4J(\boldzero)c_{\boldzero}s_{\boldzero}(c_{\boldq}^{2}+s_{\boldq}^{2})
  \sim 2J(\boldzero)-2\sqrt{\frac{J(\boldzero)}{2K}}\sqrt{\frac{K}{2J(\boldzero)}}
  =J(\boldzero),\\
  &-8J(\boldq)c_{\boldq}s_{\boldq}c_{\boldzero}s_{\boldzero}
  +4J(\boldq)c_{\boldq}s_{\boldq}(c_{\boldzero}^{2}+s_{\boldzero}^{2})
  \sim 2J(\boldzero)\sqrt{\frac{J(\boldzero)}{2K}}\sqrt{\frac{K}{2J(\boldzero)}}
  =J(\boldzero),\\  
  &8J(\boldzero)(c_{\boldzero}^{2}+s_{\boldzero}^{2}-c_{\boldzero}s_{\boldzero})
  (c_{\boldzero}^{2}+s_{\boldzero}^{2}-2c_{\boldzero}s_{\boldzero})
  \sim 8J(\boldzero)\frac{1}{2}J(\boldzero)\frac{1}{\sqrt{2J(\boldzero)K}}
  \sqrt{\frac{K}{2J(\boldzero)}}
  =2J(\boldzero),\\
  &-4J(\boldzero)
  [(c_{\boldq}^{2}c_{\boldzero}^{2}+s_{\boldq}^{2}s_{\boldzero}^{2})
    -c_{\boldzero}s_{\boldzero}(c_{\boldq}^{2}+s_{\boldq}^{2})]
  \sim -2J(\boldzero)-2J(\boldzero)\sqrt{\frac{J(\boldzero)}{2K}}\sqrt{\frac{K}{2J(\boldzero)}}
  =-3J(\boldzero),\\
  &4J(\boldq)c_{\boldq}s_{\boldq}
  (c_{\boldzero}^{2}+s_{\boldzero}^{2}-2c_{\boldzero}s_{\boldzero})
  \sim 2J(\boldzero)\frac{J(\boldzero)}{\sqrt{2J(\boldzero)K}}\sqrt{\frac{K}{2J(\boldzero)}}
  =J(\boldzero),\\
  &-4(c_{\boldzero}^{2}+s_{\boldzero}^{2}-2c_{\boldzero}s_{\boldzero})
  [J(\boldq)(c_{\boldq}^{2}+s_{\boldq}^{2})-2J(\boldzero)c_{\boldq}s_{\boldq}]
  \sim -4\sqrt{\frac{K}{2J(\boldzero)}}\frac{J(\boldzero)K}{\sqrt{2J(\boldzero)K}}
  \approx 0,\\
  &-4(c_{\boldzero}^{2}+s_{\boldzero}^{2}-2c_{\boldzero}s_{\boldzero})
  [8J(\boldzero)c_{\boldzero}s_{\boldzero}+J(\boldzero)(c_{\boldq}^{2}+s_{\boldq}^{2})
    -2J(\boldq)c_{\boldq}s_{\boldq}]
  \sim -16J(\boldzero)\sqrt{\frac{K}{2J(\boldzero)}}\sqrt{\frac{J(\boldzero)}{2K}}
  =-8J(\boldzero).\label{eq:Last}
\end{align}
(We have considered only the terms of order $J$ and neglected the higher-order terms,
such as the term of order $K$, 
because we have considered the case for $J\gg K$.) 
Combining Eqs. (\ref{eq:First}){--}(\ref{eq:Last})
with Eqs. (\ref{eq:Gam1}){--}(\ref{eq:Gam4}),
we have
\begin{align}
  &\Gamma_{\textrm{intra}}^{(0)}(\boldq)
  \sim -2J(\boldzero)+J(\boldzero)+J(\boldzero)=0,\\
  &\Gamma_{\textrm{inter}1}^{(0)}(\boldq)
  \sim 2J(\boldzero)-3J(\boldzero)+J(\boldzero)=0,\\
  &\Gamma_{\textrm{inter}2}^{(0)}(\boldq)\sim 0,\\
  &\Gamma_{\textrm{inter}3}^{(0)}(\boldq)
  \sim 8J(\boldzero)-8J(\boldzero)=0.
\end{align}
Therefore
we find that
the coefficients of
the mean-field interaction terms which arise from
the terms of $H_{\textrm{int}}$ for $\boldq=\boldq^{\prime}=\boldzero$,
for $\boldq=\boldzero$, $\boldq^{\prime}\neq\boldzero$,
and for $\boldq\neq\boldzero$, $\boldq^{\prime}=\boldzero$
become zero within the leading order.
Note that
this property holds even for $h\neq 0$
because $\cosh2\theta_{\boldq}$ and $\sinh2\theta_{\boldq}$
are independent of $h$ [see Eqs. (\ref{eq:cosh2}) and (\ref{eq:sinh2})]. 

\section{Derivation of Eqs. (\ref{eq:Hintra}) and (\ref{eq:Hinter})}

We derive Eqs. (\ref{eq:Hintra}) and (\ref{eq:Hinter}),
i.e., the mean-field interaction terms
which arise from the terms of $H_{\textrm{int}}=H_{\textrm{intra}}+H_{\textrm{inter}}$
for $\boldq\neq\boldzero$, $\boldq^{\prime}\neq\boldzero$. 
In this derivation
we replace two of the four operators for each term in Eqs. (\ref{eq:V1}){--}(\ref{eq:V5})
by the corresponding expectation value; 
for $\boldq\neq\boldzero$, $\boldq^{\prime}\neq\boldzero$,
the expectation values of two operators satisfy 
$\langle \alpha_{\boldq}^{\dagger}\alpha_{\boldq^{\prime}} \rangle=\delta_{\boldq,\boldq^{\prime}}n_{\boldq\alpha}$,
$\langle \beta_{\boldq}^{\dagger}\beta_{\boldq^{\prime}} \rangle=\delta_{\boldq,\boldq^{\prime}}n_{\boldq\beta}$, 
$\langle \alpha_{\boldq}\beta_{\boldq^{\prime}} \rangle=0$, and 
$\langle \alpha_{\boldq}^{\dagger}\beta_{\boldq^{\prime}}^{\dagger} \rangle=0$. 
Thus, for example,
the leading terms which arise from the first term of $V_{1}$ in Eq. (\ref{eq:V1})
for $\boldq\neq\boldzero$, $\boldq^{\prime}\neq\boldzero$ become
\begin{align}
  -\frac{2}{N}\sum\limits_{\boldq,\boldq^{\prime}\neq\boldzero}
  J(\boldzero)c_{\boldq}^{2}s_{\boldq^{\prime}}^{2}
  (n_{\boldq\alpha}\alpha_{\boldq^{\prime}}^{\dagger}\alpha_{\boldq^{\prime}}
  +n_{\boldq^{\prime}\alpha}\alpha_{\boldq}^{\dagger}\alpha_{\boldq}).
\end{align}
Here we have
considered only the terms proportional to $n_{\boldq\alpha}/N$ or $n_{\boldq^{\prime}\alpha}/N$
and neglected the terms proportional to $1/N$
because $n_{\boldq\alpha}\gg 1$ for small $\boldq$.
We can similarly derive the leading terms
which arise from the other terms. 
After some calculation
we can write the mean-field interaction terms
which arise from the terms of $H_{\alpha\alpha\alpha\alpha}$, $H_{\beta\beta\beta\beta}$,
$H_{\alpha\alpha\beta\beta}$, $H_{\alpha\alpha\alpha\beta}$,
and $H_{\alpha\beta\beta\beta}$ for $\boldq\neq\boldzero$, $\boldq^{\prime}\neq\boldzero$ as follows:
\begin{align}
  &H_{\alpha\alpha\alpha\alpha}
  =\frac{1}{N}\sum\limits_{\boldq,\boldq^{\prime}\neq \boldzero}
  \Gamma_{\textrm{intra}1}(\boldq,\boldq^{\prime})n_{\boldq^{\prime}\alpha}
  \alpha_{\boldq}^{\dagger}\alpha_{\boldq},\label{eq:Hintra1-MFA}\\
  &H_{\beta\beta\beta\beta}
  =\frac{1}{N}\sum\limits_{\boldq,\boldq^{\prime}\neq \boldzero}
  \Gamma_{\textrm{intra}1}(\boldq,\boldq^{\prime})n_{\boldq^{\prime}\beta}
  \beta_{\boldq}^{\dagger}\beta_{\boldq},\label{eq:Hintra2-MFA}\\
  &H_{\alpha\alpha\beta\beta}
  =\frac{1}{N}\sum\limits_{\boldq,\boldq^{\prime}\neq \boldzero}
  \Gamma_{\textrm{inter}1}(\boldq,\boldq^{\prime})
  (n_{\boldq^{\prime}\beta}\alpha_{\boldq}^{\dagger}\alpha_{\boldq}
  +n_{\boldq^{\prime}\alpha}\beta_{\boldq}^{\dagger}\beta_{\boldq})
  +\frac{1}{N}\sum\limits_{\boldq\neq \boldzero}
  \Gamma_{\textrm{inter}2}(\boldq)
  (n_{\boldq\beta}\alpha_{\boldq}^{\dagger}\alpha_{\boldq}
  +n_{\boldq\alpha}\beta_{\boldq}^{\dagger}\beta_{\boldq}),\label{eq:Hinter1-MFA}\\
  &H_{\alpha\alpha\alpha\beta}
  =\frac{1}{N}\sum\limits_{\boldq,\boldq^{\prime}\neq \boldzero}
  \Gamma_{\textrm{inter}3}(\boldq,\boldq^{\prime})
  n_{\boldq^{\prime}\alpha}
  (\alpha_{\boldq}\beta_{\boldq}+\alpha_{\boldq}^{\dagger}\beta_{\boldq}^{\dagger})
  +\frac{1}{N}\sum\limits_{\boldq\neq \boldzero}
  \Gamma_{\textrm{inter}4}(\boldq)
  n_{\boldq\alpha}
  (\alpha_{\boldq}\beta_{\boldq}+\alpha_{\boldq}^{\dagger}\beta_{\boldq}^{\dagger}),\label{eq:Hinter2-MFA}\\
  &H_{\alpha\beta\beta\beta}
  =\frac{1}{N}\sum\limits_{\boldq,\boldq^{\prime}\neq \boldzero}
  \Gamma_{\textrm{inter}3}(\boldq,\boldq^{\prime})
  n_{\boldq^{\prime}\beta}
  (\alpha_{\boldq}\beta_{\boldq}+\alpha_{\boldq}^{\dagger}\beta_{\boldq}^{\dagger})
  +\frac{1}{N}\sum\limits_{\boldq\neq \boldzero}
  \Gamma_{\textrm{inter}4}(\boldq)
  n_{\boldq\beta}
  (\alpha_{\boldq}\beta_{\boldq}+\alpha_{\boldq}^{\dagger}\beta_{\boldq}^{\dagger}),\label{eq:Hinter3-MFA}
\end{align}
where we have used Eqs. (\ref{eq:Gam-intra1}){--}(\ref{eq:Gam-inter4}).
Combining Eqs. (\ref{eq:Hintra1-MFA}){--}(\ref{eq:Hinter3-MFA})
with $H_{\textrm{intra}}=H_{\alpha\alpha\alpha\alpha}+H_{\beta\beta\beta\beta}$
and $H_{\textrm{inter}}=H_{\alpha\alpha\beta\beta}+H_{\alpha\alpha\alpha\beta}+H_{\alpha\beta\beta\beta}$,
we obtain Eqs. (\ref{eq:Hintra}) and (\ref{eq:Hinter}).

\section{Derivation of Eqs. (\ref{eq:Gm-intra1}){--}(\ref{eq:Gm-inter4})}

We derive Eqs. (\ref{eq:Gm-intra1}){--}(\ref{eq:Gm-inter4})
by estimating the leading-order terms of Eqs. (\ref{eq:Gam-intra1}){--}(\ref{eq:Gam-inter4})
in the limit $|\boldq|, |\boldq^{\prime}|\rightarrow 0$. 
We can estimate Eqs. (\ref{eq:Gam-intra1}){--}(\ref{eq:Gam-inter4}),
i.e., the coefficients of Eqs. (\ref{eq:Hintra}) and (\ref{eq:Hinter}), 
in the limit $|\boldq|, |\boldq^{\prime}|\rightarrow 0$ 
in the same way as
the estimation of $\Gamma_{\textrm{intra}}^{(0)}(\boldq)$,
$\Gamma_{\textrm{inter}1}^{(0)}(\boldq)$,
$\Gamma_{\textrm{inter}2}^{(0)}(\boldq)$, and 
$\Gamma_{\textrm{inter}3}^{(0)}(\boldq)$ [for the latter estimation see Appendix C,
in particular, the derivation of Eqs. (\ref{eq:First}){--}(\ref{eq:Last})].
As a result,
we obtain
\begin{align}
  \Gamma_{\textrm{intra}1}(\boldq,\boldq^{\prime})
  =&2J(\boldzero)-2J(\boldzero)\cosh 2\theta_{\boldq}\cosh 2\theta_{\boldq^{\prime}}
  -2J(\boldq-\boldq^{\prime})\sinh 2\theta_{\boldq}\sinh 2\theta_{\boldq^{\prime}}
  +2J(\boldq)\sinh 2\theta_{\boldq}\cosh 2\theta_{\boldq^{\prime}}\notag\\
  &+2J(\boldq^{\prime})\cosh 2\theta_{\boldq}\sinh 2\theta_{\boldq^{\prime}}\notag\\
  \sim & 2J(\boldzero)
       [1-\frac{J(\boldzero)^{2}+2J(\boldzero)K}{2J(\boldzero)K}
         -\frac{J(\boldzero)^{2}}{2J(\boldzero)K}
         +\frac{J(\boldzero)^{2}+J(\boldzero)K}{2J(\boldzero)K}
         +\frac{J(\boldzero)^{2}+J(\boldzero)K}{2J(\boldzero)K}]\notag\\
  =&2J(\boldzero),\label{eq:Gam-intra1-estimate}\\
  \Gamma_{\textrm{inter}1}(\boldq,\boldq^{\prime})
  =&-4J(\boldzero)+\Gamma_{\textrm{intra}1}(\boldq,\boldq^{\prime})\notag\\
  \sim & -2J(\boldzero),\label{eq:Gam-inter1-estimate}\\
  \Gamma_{\textrm{inter}2}(\boldq)
  =&-4\cosh 2\theta_{\boldq}[J(\boldzero)\cosh 2\theta_{\boldq}-J(\boldq)\sinh 2\theta_{\boldq}]\notag\\
  \sim & -4J(\boldzero)+2J(\boldzero)\notag\\
  =&-2J(\boldzero),\label{eq:Gam-inter2-estimate}\\
  \Gamma_{\textrm{inter}3}(\boldq,\boldq^{\prime})
  =&2J(\boldzero)\sinh 2\theta_{\boldq}\cosh 2\theta_{\boldq^{\prime}}
  +2J(\boldq-\boldq^{\prime})\sinh 2\theta_{\boldq^{\prime}}\cosh 2\theta_{\boldq}
  -2J(\boldq)\cosh 2\theta_{\boldq}\cosh 2\theta_{\boldq^{\prime}}\notag\\
  &-2J(\boldq^{\prime})\sinh 2\theta_{\boldq^{\prime}}\sinh 2\theta_{\boldq}\notag\\
  \sim & [J(\boldzero)\frac{J(\boldzero)}{K}+J(\boldzero)]
  +[J(\boldzero)\frac{J(\boldzero)}{K}+J(\boldzero)]
  -[J(\boldzero)\frac{J(\boldzero)}{K}+2J(\boldzero)]
  -J(\boldzero)\frac{J(\boldzero)}{K}\notag\\
  =&0,\label{eq:Gam-inter3-estimate}\\
  \Gamma_{\textrm{inter}4}(\boldq)
  =&4J(\boldzero)\sinh 2\theta_{\boldq}\cosh 2\theta_{\boldq}
  -2J(\boldq)(\cosh 2\theta_{\boldq})^{2}
  -2J(\boldq)(\sinh 2\theta_{\boldq})^{2}\notag\\
  \sim & [2J(\boldzero)\frac{J(\boldzero)}{K}+2J(\boldzero)]
  -[J(\boldzero)\frac{J(\boldzero)}{K}+2J(\boldzero)]
  -J(\boldzero)\frac{J(\boldzero)}{K}\notag\\
  =&0.\label{eq:Gam-inter4-estimate}
\end{align}
Note that 
Eqs. (\ref{eq:Hinter2-MFA}) and (\ref{eq:Hinter3-MFA})
with Eqs. (\ref{eq:Gam-inter3-estimate}) and (\ref{eq:Gam-inter4-estimate})
show that
the contributions from $H_{\alpha\alpha\alpha\beta}$ and $H_{\alpha\beta\beta\beta}$,
corresponding to the mean-field interaction terms including $\alpha_{\boldq}\beta_{\boldq}$
and $\alpha_{\boldq}^{\dagger}\beta_{\boldq}^{\dagger}$,
are zero within the leading order. 

\end{widetext}


\begin{references}
\bibitem{Leggett-review}
  A. J. Leggett, 
  Rev. Mod. Phys. \textbf{73}, 307 (2001).

\bibitem{Einstein}
  A. Einstein,
  Sitzungsberichte der Preussischen Akademie der Wissenschaften,
  Physikalisch-mathematische Klasse 261 (1924);
  A. Einstein, 
  Sitzungsberichte der Preussischen Akademie der Wissenschaften,
  Physikalisch-mathematische Klasse 3 (1925).

\bibitem{BEC-text}
  C. J. Pethick and H. Smith, 
  \textit{Bose-Einstein Condensation in Dilute Gases} 
  (Cambridge University Press, Cambridge, England, 2002).

\bibitem{FW}
  A. L. Fetter and J. D. Walecka, 
  \textit{Quantum Theory of Many-Particle Systems} 
  (Dover Publications, Inc., New York, 2003).

\bibitem{MagBEC-JETP}
  Yu. D. Kalafati and V. L. Safonov,
  Pis'ma Zh. Eksp. Teor. Fiz. \textbf{50}, 135 (1989)
  [JETP Lett. \textbf{50}, 149 (1989)].

\bibitem{MagBEC-Nature}
  S. O. Demokritov, V. E. Demidov, O. Dzyapko, G. A. Melkov, 
  A. A. Serga, B. Hillebrands, and A. N. Slavin,
  Nature (London) \textbf{443}, 430 (2006).

\bibitem{MagBEC-He}
  Yu. M. Bunkov and G. E. Volovik,
  Phys. Rev. Lett. \textbf{98}, 265302 (2007).

\bibitem{MagBEC-AF1}
  Yu. M. Bunkov, E. M. Alakshin, R. R. Gazizulin, A. V. Klochkov,
  V. V. Kuzmin, T. R. Safin, and M. S. Tagirov,
  JETP Lett. \textbf{94}, 68 (2011); 
  E. M. Alakshin, Yu. M. Bunkov, R. R. Gazizulin, A. V. Klochkov,
  V. V. Kuzmin, A. S. Nizamutdinov, T. R. Safin, and M. S. Tagirov, 
  J. Phys. Conf. Ser. \textbf{324}, 012006 (2011).

\bibitem{MagBEC-AF2}
  Yu. M. Bunkov, E. M. Alakshin, R. R. Gazizulin, A. V. Klochkov,
  V. V. Kuzmin, V. S. L’vov, and M. S. Tagirov,
  Phys. Rev. Lett. \textbf{108}, 177002 (2012).

\bibitem{MagBEC-AF3}
  E. L. Fj{\ae}rbu, N. Rohling, and A. Brataas, 
  Phys. Rev. B \textbf{95}, 144408 (2017).  

\bibitem{NA-Ferri}
  N. Arakawa,
  Phys. Rev. Lett. \textbf{121}, 187202 (2018). 

\bibitem{Anderson-AF-mag}
  P. W. Anderson, 
  Phys. Rev. \textbf{86}, 694 (1952).

  \bibitem{Nakamura}
  T. Nakamura and M. Bloch, 
  Phys. Rev. \textbf{132}, 2528 (1963).

\bibitem{HP}
  T. Holstein and H. Primakoff, 
  Phys. Rev. \textbf{58}, 1098 (1940).
  
\bibitem{MagBEC-theory}
  I. S. Tupitsyn, P. C. E. Stamp, and A. L. Burin, 
  Phys. Rev. Lett. \textbf{100}, 257202 (2008).



\bibitem{Yosida}
  K. Yosida, 
  Prog. Theor. Phys. \textbf{6}, 691 (1951).

\bibitem{AF-exp}
  C. Trapp and J. W. Stout, 
  Phys. Rev. Lett. \textbf{10}, 157 (1963).

\bibitem{Kubo-AF-mag}
  R. Kubo, 
  Phys. Rev. \textbf{15}, 568 (1952).

\bibitem{Oguchi}
  T. Oguchi, 
  Phys. Rev. \textbf{117}, 117 (1960); 
  T. Oguchi and A. Honma, 
  J. Phys. Soc. Jpn. \textbf{16}, 79 (1961).

    
\bibitem{remark}
  Using $n[\epsilon(\boldq)\pm h]\approx n[\epsilon(\boldq)]\pm h n^{\prime}[\epsilon(\boldq)]$,
  where $n^{\prime}(\epsilon)=\partial n(\epsilon)/\partial \epsilon < 0$,
  we obtain
  $n[\epsilon_{\alpha}(\boldq^{\prime})]
  -(1+\delta_{\boldq^{\prime},\boldq})n[\epsilon_{\beta}(\boldq^{\prime})]
  \approx -\delta_{\boldq^{\prime},\boldq}n[\epsilon(\boldq^{\prime})]+(2+\delta_{\boldq^{\prime},\boldq}) h n^{\prime}[\epsilon(\boldq^{\prime})]$. 
  This shows the existence of the critical magnetic field.  


 

\bibitem{AF-Mag}
  S. Takei, B. I. Halperin, A. Yacoby, and Y. Tserkovnyak,
  Phys. Rev. B \textbf{90}, 094408 (2014). 

\bibitem{Review1}
  T. Jungwirth, X. Marti, P. Wadley, and J. Wunderlich,
  Nat. Nanotechnol. \textbf{11}, 231 (2016).
  
\bibitem{Review2}
  V. Baltz, A. Manchon, M. Tsoi, T. Moriyama, T. Ono, and Y. Tserkovnyak,
  Rev. Mod. Phys. \textbf{90}, 015005 (2018).

  \bibitem{Bogoliubov}
  N. N. Bogoliubov, 
  Izv. Akad. Nauk SSSR, Ser. Fiz. \textbf{11}, 77 (1947)
  [J. Phys. (USSR) \textbf{11}, 23 (1947)]. 
  
\bibitem{AGD}
  A. A. Abrikosov, L. P. Gor'kov and I. E. Dzyaloshinski, 
  \textit{Methods of Quantum Field Theory in Statistical Physics}
  (Dover Publications, Inc., New York, 1963).
\end{references}
\end{document}